\documentclass[lettersize,journal]{IEEEtran}
\usepackage{amsmath,amssymb,amsfonts,amsthm}
\usepackage{algorithmic}
\usepackage{algorithm}
\usepackage{array}
\usepackage{arydshln}
\usepackage[caption=false,font=normalsize,labelfont=sf,textfont=sf]{subfig}
\usepackage{textcomp}
\usepackage{stfloats}
\usepackage{url}
\usepackage{color}
\usepackage{verbatim}
\usepackage{graphicx}
\usepackage{cite}
\newtheorem{definition}{{Definition}}
\usepackage{hyperref}
\usepackage[normalem]{ulem}
\usepackage{multirow}
\usepackage{enumitem}
\usepackage{float}
\usepackage{caption}
\usepackage{subcaption}
\usepackage{makecell}

\usepackage{xspace}

\newcommand{\eg}{e.g.,\xspace}
\newcommand{\ie}{i.e.,\xspace}

\newcommand{\model}{\emph{FlowID}\xspace}

\begin{document}
\title{Multi-view Correlation-aware Network Traffic Detection on Flow Hypergraph}

\author{Jiajun Zhou, Wentao Fu, Hao Song, Shanqing Yu, Qi Xuan,~\IEEEmembership{Senior Member,~IEEE}

\IEEEcompsocitemizethanks{
\IEEEcompsocthanksitem This work was supported in part by National Natural Science Foundation of China under Grant 62503423, in part by the Key Research and Development Program of Zhejiang under Grants 2022C01018 and 2024C01025, in part by the National Natural Science Foundation of China under Grant U21B2001, and in part by the Baima Lake Laboratory Joint Fund of Zhejiang Provincial Natural Science Foundation of China under Grant LBMHZ25F020002.
\emph{(Corresponding author: Shanqing Yu.)}
\IEEEcompsocthanksitem Jiajun Zhou is with the Institute of Cyberspace Security, Zhejiang University of Technology, Hangzhou 310023, China, with the Binjiang Institute of Artificial Intelligence, ZJUT, Hangzhou 310056, China (e-mail: jjzhou@zjut.edu.cn).
\IEEEcompsocthanksitem Wentao Fu, Hao Song, Shanqing Yu and Qi Xuan are with the Institute of Cyberspace Security, College of Information Engineering, Zhejiang University of Technology, Hangzhou 310023, China, with the Binjiang Institute of Artificial Intelligence, ZJUT, Hangzhou 310056, China (e-mail:\{fuwt, songhao, yushanqing, xuanqi\}@zjut.edu.cn).
\IEEEcompsocthanksitem Both W. Fu and H. Song make equal contributions to this work.
}}



\maketitle

\begin{abstract}


As the Internet rapidly expands, the increasing complexity and diversity of network activities pose significant challenges to effective network governance and security regulation. Network traffic, which serves as a crucial data carrier of network activities, has become indispensable in this process. Network traffic detection aims to monitor, analyze, and evaluate the data flows transmitted across the network to ensure network security and optimize performance. However, existing network traffic detection methods generally suffer from several limitations: 1) a narrow focus on characterizing traffic features from a single perspective; 2) insufficient exploration of discriminative features for different traffic; 3) poor generalization to different traffic scenarios. To address these issues, we propose a multi-view correlation-aware framework named \model for network traffic detection. \model captures multi-view traffic features via temporal and interaction awareness, while a hypergraph encoder further explores higher-order relationships between flows. To overcome the challenges of data imbalance and label scarcity, we design a dual-contrastive proxy task, enhancing the framework's ability to differentiate between various traffic flows through flow-to-flow and group-to-group contrast. Extensive experiments on five real-world datasets demonstrate that \model significantly outperforms existing methods in accuracy, robustness, and generalization across diverse network scenarios, particularly in detecting malicious traffic.

\end{abstract}


\section{Introduction}
\IEEEPARstart{A}{s} the global Internet of Everything trend accelerates, the boundaries of the internet are rapidly expanding, fostering closer connections among individuals. Simultaneously, the widespread adoption of IoT devices, such as smart homes, smart city infrastructure, and autonomous driving, has also made human-object interactions more intelligent. While this trend facilitates information exchange and resource sharing, it also presents unprecedented challenges in network governance and security regulation.
Network traffic, as a critical carrier that records and reflects the activities of networks and their users, has become an indispensable object in network governance and regulation. Network traffic detection aims to analyze network traffic data in order to determine the underlying network activities and behaviors, such as email communication, file transfer, streaming services, port scanning, malware propagation, and intrusion attacks. Accurately identifying the characteristics and patterns exhibited by network traffic plays a critical role in optimizing resource allocation within the network infrastructure, proactively preventing and responding to cyber-attacks, as well as ensuring the stable operation of information systems.



Early detection methods typically focus on explicit elements such as packet size~\cite{lang2003synthetic}, packet timestamp~\cite{paxson1994empirically}, and port number~\cite{zhang2014robust}. These methods involve manually designing features and training machine learning (ML) models to extract and classify traffic characteristics~\cite{nguyen2008survey,kotpalliwar2015classification,kokila2014ddos}. 
However, as network activities have become increasingly frequent and diverse, effectively processing and identifying large volumes of unstructured traffic data has become more challenging. ML-based methods heavily rely on feature engineering, which is not only time-consuming and labor-intensive, but also struggles to adapt to the rapid changes in cyberspace situation. This issue is particularly pronounced in scenarios involving complex cyber-attacks and encrypted communications, where detection efficiency and accuracy are severely limited.

The emergence of deep learning (DL) techniques has provided new perspectives for network traffic analysis~\cite{kwon2019survey,liu2019machine}. Deep neural networks possess the ability to automatically acquire implicit features from traffic data, thereby significantly enhancing the accuracy and efficiency of traffic detection. 
For example, GraphDDoS~\cite{li2022graphddos} and GraphDApp~\cite{shen2021accurate} model network traffic as Traffic Interaction Graphs (TIGs), where packets are represented as nodes and packet direction and length serve as edge attributes. These approaches leverage graph neural networks (GNNs) to classify TIGs for traffic analysis. Other works, such as SmartDetector~\cite{shen2025robust}, characterize individual packet features by constructing a semantic attribute matrix. Additionally, methods like ATVITSC~\cite{liu2024atvitsc} and FS-MTC~\cite{zhang2024enhanced} model network traffic as grayscale images and apply convolutional neural networks (CNNs) to extract features from these representations for malicious traffic detection.
Nevertheless, these DL-based approaches also exhibit certain limitations: 
1) \textbf{Limited characterization of individual traffic features.} 
Existing models typically focus on single aspects when extracting traffic features, such as statistical properties, temporal features, and packet interaction characteristics, neglecting multi-view characterization of network traffic;
2) \textbf{Insufficient exploration of inter-traffic relationships.} 
Most existing methods analyze network traffic characteristics from an isolated perspective, overlooking the fact that network behavior may be continuous, thereby failing to capture the implicit correlations and dependencies between traffic; 
3) \textbf{Poor generalization to different network scenarios.}
Current network traffic detection methods are often trained using data from specific scenarios, making them less adaptable to complex public networks containing mixed encrypted and unencrypted traffic. Moreover, due to the inherent imbalance in the distribution of traffic categories (e.g., malicious traffic being less frequent than normal traffic), these models demonstrate limited generalization to diverse Internet or IoT traffic scenarios.



With the increasing frequency and diversity of network activities, effectively processing and identifying large-scale, unstructured, and dynamically changing traffic data has become a formidable challenge. To address these, we propose a multi-view correlation-aware framework for network traffic detection, named \model. 
Specifically, we first design a multi-view feature extractor to capture the features of individual traffic flows in complex scenarios. Based on these features, we further construct a flow hypergraph to explicitly present the high-order correlations among individual traffic flows and use a hypergraph encoder to learn the features of both flows and flow groups.
To address the limitations in label availability for traffic data, we design dual-contrastive self-supervised proxy tasks: flow-to-flow contrast and group-to-group contrast, to further enhance the discriminative power of our hypergraph encoder across different traffic flow types.
Extensive experiments on five real-world network traffic datasets demonstrate that our framework outperforms existing methods in network traffic detection, as well as in robustness against network interference. Moreover, in light of the significant class imbalance and label scarcity issues in traffic data, our framework, under the optimization of dual-contrastive constraints, exhibits strong generalization capability, enabling effective identification of various traffic flows across different network scenarios.
The main contributions are summarized as follows:
\begin{itemize}[leftmargin=15pt]
    \item \textbf{Powerful feature characterization:} We capture the temporal and interaction information of packets within individual flows through multi-view feature extraction, and further integrate higher-order correlations between flows using hypergraph representation learning, enabling a comprehensive characterization of traffic features.
    \item \textbf{Generalization:} We design dual-contrastive proxy tasks to optimize our framework, addressing the challenges posed by imbalanced class distribution and label scarcity.
    \item \textbf{State-of-the-art performance:} Extensive experiments on network traffic datasets demonstrate the effectiveness and superiority of our framework in traffic detection.
\end{itemize}


\section{Related Work} \label{sec: related-work}
\subsection{Network Traffic Detection}
Network traffic detection involves analyzing and processing network packets to determine their application categories, service types, and potential security risks. Traditional machine learning approaches based on statistical principles often rely on manual feature extraction, where domain experts transform raw traffic data into a format suitable for algorithmic classification.
Moore et al.~\cite{moore2005toward} extract statistical features from traffic data generated by different applications and trained a Bayesian classifier for traffic identification.
AppScanner~\cite{taylor2016appscanner} utilizes Random Forest (RF) model to distinguish mobile applications based on statistical features extracted from traffic data.
FlowPrint~\cite{van2020flowprint} combines destination-based clustering, browser isolation, and pattern recognition in a semi-supervised framework for application fingerprinting.
Saber et al.~\cite{saber2018encrypted} propose integrating Principal Component Analysis (PCA) with Support Vector Machines (SVM) to identify traffic, focusing on time-based traffic features.
While these traditional methods can be effective in certain scenarios, they heavily depend on domain-specific feature engineering.
\begin{figure*}[!htb]
	\centering
	\includegraphics[width=\textwidth]{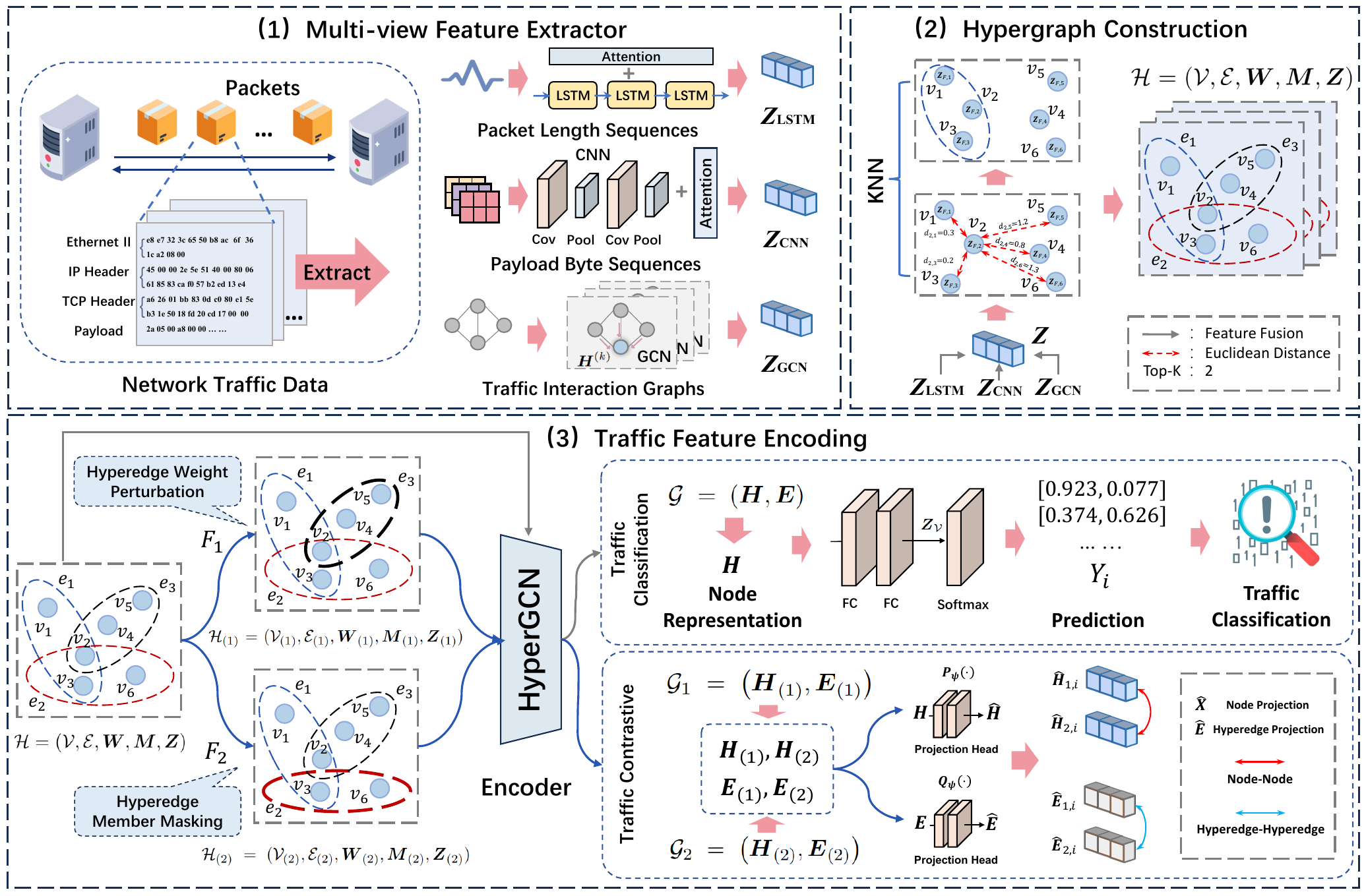}    
	\caption{The architecture of \model: 
     1) Multi-view traffic feature extraction via temporal feature awareness and interaction feature awareness; 
     2) Traffic hypergraph construction to capture higher-order flow relationships; 
     3) Traffic hypergraph learning using hypergraph convolution and double contrastive learning.}
	\label{fig: Framework}
\end{figure*}

With the advancement of deep learning, deep neural networks have exhibited powerful capabilities in extracting features, thereby enabling them to autonomously acquire intricate patterns from raw data. Consequently, network traffic identification methods based on deep learning have garnered substantial attention.
FS-Net~\cite{liu2019fs} uses packet length sequences as inputs, employing bidirectional GRUs for feature encoding and introducing a reconstruction mechanism in an Autoencoder to ensure the effectiveness of the learned features.
MTT~\cite{zheng2022mtt} employs a multi-task Transformer trained on truncated packet byte sequences to analyze traffic characteristics in a supervised manner. 
NetMamba~\cite{wang2024netmamba} introduces a novel state-space pre-trained model that replaces the Transformer architecture, offering improved efficiency in traffic sequence modeling.
However, these methods generally overlook the interaction between entities within the network.

Graph-based approaches address this gap by treating traffic packets as nodes and packet transmission relationships as edges, constructing traffic packet interaction graphs.
Graph representation learning techniques, such as graph neural networks (GNNs)~\cite{zhou2025ncgnn}, are then applied to extract both network topology and traffic interaction information~\cite{huoh2022flow,barsellotti2023ftg}.
Han et al.~\cite{han2024intrusion} introduce Graph Integration Theory (GIT), which enhances intrusion detection capabilities for encrypted traffic in IoT by utilizing only the sequence of packet lengths as a single feature. 
DE-GNN~\cite{han2024gnn} employs dual embedding layers to encode both packet headers and payload bytes, and uses PacketCNN to derive packet-level representations of traffic, facilitating the construction of interaction graphs.

\subsection{Hypergraph Neural Network}
Graph neural networks (GNNs) have become essential for analyzing non-Euclidean data, with applications spanning recommendation systems~\cite{ying2018graph}, community discovery~\cite{zhou2021robustecd,chen2019supervised}, and blockchain system~\cite{jin2022heterogeneous,jin2024time,zhou2022behavior,du2023breaking,che2024across}. However, traditional GNNs are limited by their focus on pairwise relationships, which is insufficient for modeling the complex interactions seen in many real-world scenarios. Hypergraph neural networks (HGNNs) address this limitation by extending traditional edges to hyperedges, allowing for higher-order interactions between multiple nodes~\cite{feng2019hypergraph}.
Yadati et al.~\cite{yadati2019hypergcn} propose HyperGCN to address the challenges of semi-supervised classification problems on hypergraphs. Since then, a plethora of HGNN models have been successfully employed in various domains including computer vision~\cite{chen2020hypergraph} and biochemistry~\cite{wang2020next}, yielding remarkable outcomes.
Additionally, in the field of network traffic analysis, Pu et al.~\cite{pu2011hypergraph} propose a clustering strategy based on hypergraph models to effectively group open ports and users in networks, thereby reducing the manual inspection workload associated with data volume.
RHCRN~\cite{yu2023routing} employs routing hypergraph convolutional recursive networks to treat routing paths as hyperedges and explore intricate spatial correlations between routing paths and network nodes.
HRNN~\cite{yang2024hrnn} models traffic as nodes, integrating the structural characteristics of hypergraphs with the temporal processing capabilities of Recurrent Neural Networks (RNNs), effectively capturing and analyzing both spatial and temporal features in traffic, thereby facilitating intrusion detection in networks.

\section{Methodology}\label{sec: model}

In this section, we provide the details of our proposed framework \model, as schematically depicted in Fig.~\ref{fig: Framework}. The framework is composed of three main components: 
1) a multi-view traffic feature extraction module, which is used to extract temporal and interaction features from the traffic data; 
2) a flow hypergraph construction module, which explicitly represents the relationships between flows through the hypergraph structure, capturing higher-order relationships in complex network environments;
3) a flow hypergraph learning module, which further enhances the robustness and generalization of traffic feature representations through hypergraph convolutional networks and dual-contrastive learning.

\subsection{Multi-view Traffic Feature Extraction}\label{sec: mv}
This module models individual flows from multiple perspectives, aiming to capture various features of flows at the micro level. Specifically, we primarily focus on the packet size, transmission direction, payload patterns, and packet interactions, forming a multi-view traffic feature extraction module that integrates temporal features and interaction features.

\subsubsection{Temporal Feature Awareness}


Packets in traffic flows are the fundamental units that construct network communications. A flow consists of numerous packets, which are typically arranged in chronological order. By analyzing the temporal characteristics of the information within these packets, we can reveal behavioral patterns of network activities and identify potential anomalies.
Specifically, we first extract the transmission direction and length information of packets within a flow and arrange them in chronological order to form a \textbf{directed length sequence of packets} ($\textsf{Seq}_\text{len}$). 
In network traffic analysis, critical information is typically concentrated in the initial phase of a flow, especially during connection establishment, protocol negotiation, and initial request exchanges. Due to the high computational cost and low efficiency of processing complete flow data, we extract the first $n$ packets of the flow to construct feature sequences, for flows with fewer than $n$ packets, zero-padding is applied to maintain a consistent sequence length across different flows.
The directed length sequences extracted from all flows can be represented as a tensor $\boldsymbol{X}_\text{len} \in \mathbb{R}^{N\times n}$, where $N$ denotes the number of flows. We then employ a Long Short-Term Memory (LSTM) network~\cite{lstm} to analyze the temporal characteristics within $\textsf{Seq}_\text{len}$, and introduce an attention mechanism to aggregate features across different time steps, ultimately acquiring insights into the transmission patterns and application types associated with each flow:
\begin{equation}
    \boldsymbol{Z}_{\text{LSTM}} = \text{Attention}\left( \text{LSTM}(\boldsymbol{X}_\text{len}) \right)
\end{equation}


In network communication, the payload carries protocol-specific content or user interaction data and serves as a critical component for traffic analysis. For unencrypted traffic, the raw byte sequence of the payload directly reveals application-layer protocol characteristics, facilitating the extraction of user behavior information. In contrast, while the payload content is obfuscated in encrypted traffic, we observe that mainstream encryption protocols (\eg TLS, SSH) typically produce high-entropy data with nearly uniform byte distributions. These statistical properties can be leveraged to distinguish between encrypted and unencrypted traffic in complex network environments, providing auxiliary cues for encrypted traffic detection and classification.
Therefore, for each flow, we further extract the payload information of its first $n$ packets. Specifically, for the payload of each packet, we extract the first $m$ bytes of information to improve detection efficiency and convert them from hexadecimal to decimal, forming a \textbf{payload byte sequence} ($\textsf{Seq}_\text{byte}$). For packets with less than $m$ bytes of payload, we pad the sequence with zeros to ensure consistent sequence length. Ultimately, the payload byte sequences extracted from all traffic data can be denoted as a tensor $\boldsymbol{X}_\text{byte} \in \mathbb{R}^{N\times n\times m}$, where the payload information of all packets in each flow is encapsulated as a two-dimensional matrix $\in\mathbb{R}^{n\times m}$, with each packet's payload information stored in a row vector $\in\mathbb{R}^{m}$ of the matrix.
Subsequently, a convolutional neural network (CNN) is employed to analyze the payload information of all flows. By stacking multiple layers of convolution, pooling, and non-linear activation functions, CNN can automatically extract high-dimensional features from abstract payload data. We employ CNN to learn the spatial dependencies and contextual relationships within $\textsf{Seq}_\text{byte}$, as well as the correlations between different packet payloads. This facilitates a comprehensive understanding of the protocol and application characteristics of flows, along with the overall behavioral patterns. Additionally, we also introduce an attention mechanism to help the model focus on crucial information within the payload, thereby achieving efficient payload feature extraction.
\begin{equation}
    \boldsymbol{Z}_{\text{CNN}}=\text{Attention}(\text{CNN}(\boldsymbol{X}_\text{byte}))
\end{equation}

\subsubsection{Interaction Feature Awareness}
During temporal feature awareness, by constructing the sequence modality of flows and analyzing the sequence characteristics such as transmission direction, packet length, and payload information, we can capture fundamental temporal patterns within flows. However, relying solely on sequence analysis often provides only a limited perspective of the flows, which may fail to capture the intricate dependencies and interactions among packets.

\begin{figure}[!htb]
	\centering
	\includegraphics[width=\linewidth]{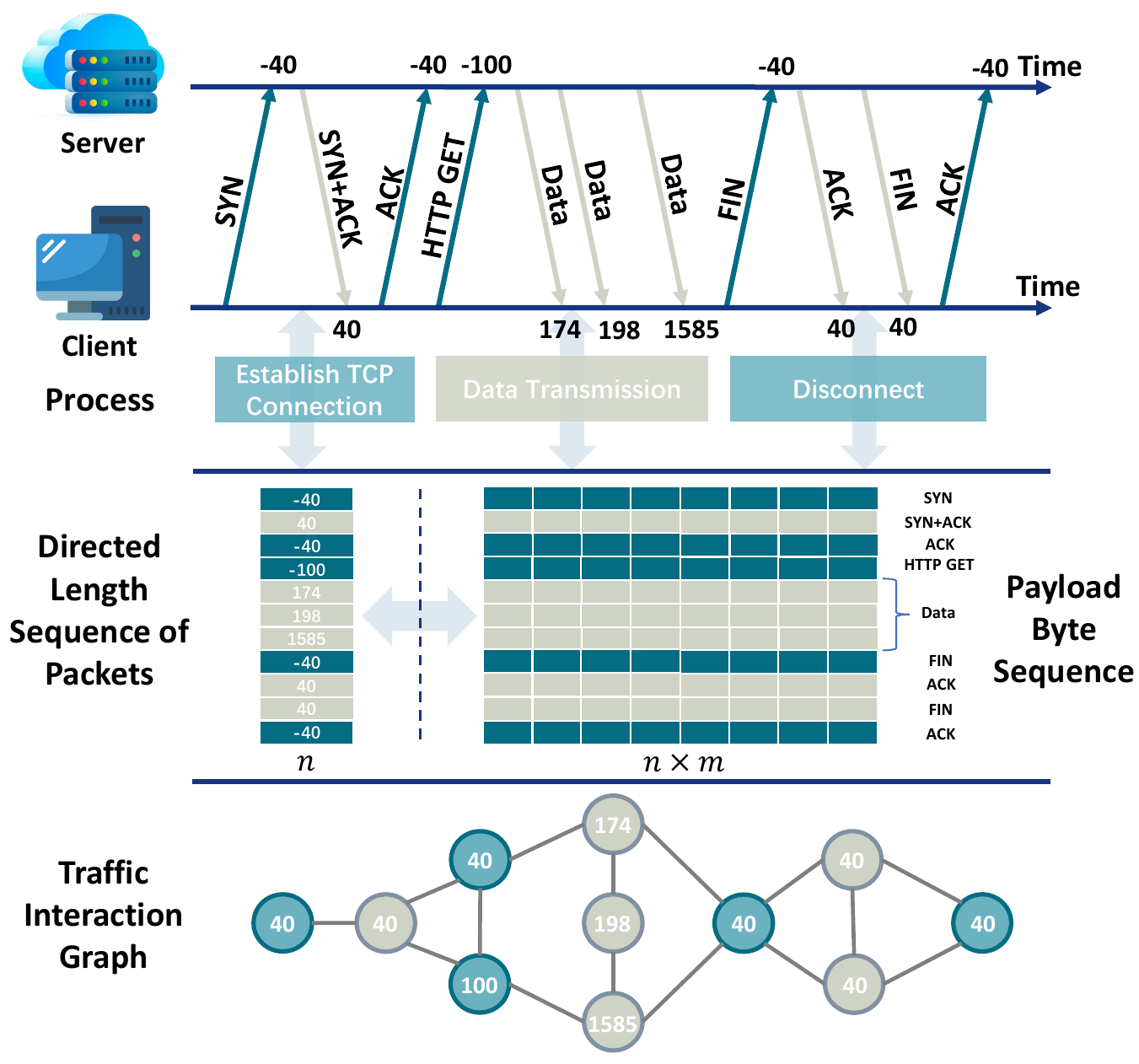}      
	\caption{{An example of traffic sequence and graph construction under HTTP request-response.}}
	\label{fig: TIG}
\end{figure}

To address this, we further utilize a Traffic Interaction Graph (TIG)~\cite{shen2021accurate} to reflect the internal behavioral mechanisms of flows. Fig.~\ref{fig: TIG} shows an example of constructing a TIG based on HTTP request-response. Specifically, we mark the direction from the client to the server as negative and the direction from the server to the client as positive. Such practice takes into account the bidirectionality of the communication process, thereby facilitating an accurate representation of network traffic behavior. We then represent each packet as a node, with its length and direction as node features, and the communication relationships between packets as edges, marking packets of different directions with different colors. Additionally, the TIG incorporates the concept of layers, grouping consecutive packets with the same direction into the same layer (represented as columns in Fig.~\ref{fig: TIG}). Furthermore, we optimize the construction of TIG. Specifically, considering the variation in the duration of different flows, we utilize the first $n$ packets from these prolonged flows to represent them and construct the corresponding TIG based on this subset. This optimization reduces computational consumption and enhances the efficiency of network traffic analysis.

We represent the TIG corresponding to a flow as $\mathcal{G}=(\mathcal{V},\mathcal{E},\boldsymbol{X})$, where $\mathcal{V}$ is the set of packets, $\mathcal{E}$ is the set of sequential relationship edges between packets, and $\boldsymbol{X} \in \mathbb{R}^{n\times 2}$ is the feature matrix composed of packet length and direction. The topological structure of the TIG can also be represented by an adjacency matrix $\boldsymbol{A} \in \mathbb{R}^{n\times n}$, where entry $\boldsymbol{A}_{ij}=1$ if there exists an edge between packet $i$ and packet $j$, and $\boldsymbol{A}_{ij}=0$ otherwise. On this basis, we further utilize a Graph Convolutional Network (GCN)~\cite{kipf2016semi} to learn the communication patterns and behavioral characteristics of flows presented by the packet interaction structures within the TIG. The GCN takes the structural and feature information of the TIG as input and updates the representations of packet nodes by propagating, aggregating, and transforming node features across the graph. Ultimately, a global mean pooling operation is employed to aggregate information from all packet nodes, yielding the representation of the entire TIG (or flow). The completed feature extraction process using two-layer GCN is presented as follows:
\begin{equation}
    \boldsymbol{Z}_\text{GCN} = \text{MeanPooling}\left( \sigma\left(\boldsymbol{\tilde{A}}_\text{n} \cdot \sigma\left(\boldsymbol{\tilde{A}}_\text{n} \boldsymbol{X} \boldsymbol{W}^{(1)}_\text{g}\right) \cdot \boldsymbol{W}^{(2)}_\text{g}\right) \right)
\end{equation}
where $\boldsymbol{W}^\text{g}$ is the weight matrix, $\sigma(\cdot)$ is the activation function, $\tilde{\boldsymbol{A}}_\text{n}=\tilde{\boldsymbol{D}}^{-1/2}\tilde{\boldsymbol{A}}\tilde{\boldsymbol{D}}^{-1/2}$ is the symmetric normalized form of $\tilde{\boldsymbol{A}}$, and $\tilde{\boldsymbol{A}}=\boldsymbol{A}+\boldsymbol{I}$ is the adjacency matrix with self-loop, $\tilde{\boldsymbol{D}}$ is the degree matrix of $\tilde{\boldsymbol{A}}$ with $\tilde{\boldsymbol{D}}_{ii} = \sum_j \tilde{\boldsymbol{A}}_{ij}$.

\subsubsection{Multi-view Feature Fusion}
{During temporal and interaction feature awareness, we obtain the multi-view features of flows. To better represent individual flows, we fuse these three types of features. Specifically, we first concatenate the two temporal features and apply multiple linear transformations and dropout operation to enhance the expressiveness and generalization of the fused features:
\begin{equation} 
     \boldsymbol{Z}_{\text{SEQ}}= \text{Linear(Dropout}(\text{Linear}(\boldsymbol{Z}_{\text{CNN}} \parallel \boldsymbol{Z}_{\text{LSTM}})))
\end{equation}
where $\parallel$ denotes the concatenation operation. 
Subsequently, we combine the interaction features with temporal features through adaptive linear interpolation, obtaining the final representation of network flows:
\begin{equation}  
    \boldsymbol{Z}_\text{mv}=\alpha \cdot \boldsymbol{Z}_\text{GCN} + (1-\alpha)\cdot \boldsymbol{Z}_{\text{SEQ}}
\end{equation}
where the fusion weight $\alpha$ is a learnable parameter. }

\subsection{Flow Hypergraph Construction}
Existing traffic detection methods typically focus solely on analyzing individual flows, neglecting the continuity of network activities and the correlations between flows. To enhance the capability of identifying traffic behaviors in complex network scenarios, we further propose constructing a flow hypergraph to explicitly represent flow groups, providing a structural foundation for mining implicit relationships among flows. A hypergraph, as a generalized graph model, allows hyperedges to connect multiple nodes, offering greater flexibility and expressiveness than traditional graph models in capturing complex higher-order relationships.
\begin{definition}
    \textnormal{(\textbf{Flow Hypergraph.})} \ 
    A flow hypergraph can be represented as $\mathcal{H}=(\mathcal{V},\mathcal{E},\boldsymbol{M},\boldsymbol{H},\boldsymbol{Z})$, where $\mathcal{V}$ is the set of nodes (flows), $\mathcal{E}$ is the set of hyperedges (flow groups), and $\boldsymbol{Z}$ is the node feature matrix. 
    The hyperedge weight matrix $\boldsymbol{M}\in\mathbb{R}^{|\mathcal{E}|\times |\mathcal{E}|}$, which is a diagonal matrix with each element $\boldsymbol{M}_{ii}$ represents the weight of the corresponding hyperedge $e_{i}$.
    The incidence matrix $\boldsymbol{H}\in\mathbb{R}^{|\mathcal{V}|\times |\mathcal{E}|}$ represents the association between nodes and hyperedges, where $\boldsymbol{H}_{ij}=1$ means that node $v_i$ is in hyperedge $e_j$, and $\boldsymbol{H}_{ij}=0$ otherwise.
    The node degree matrix is represented as $\boldsymbol{D}_v\in\mathbb{R}^{|\mathcal{V}|\times |\mathcal{V}|}$, which is a diagonal matrix with the diagonal element $\boldsymbol{D}_v[i,i]$ denoting the sum of the weights associated with hyperedges that involve the node $v_i$.
    The hyperedge degree matrix is represented as $\boldsymbol{D}_e\in\mathbb{R}^{|\mathcal{E}|\times |\mathcal{E}|}$, which is also a diagonal matrix with the diagonal element $\boldsymbol{D}_e[j,j]$ denoting the number of nodes connected by the hyperedge $e_j$.
\end{definition}


During flow hypergraph construction, each flow is treated as a node, with its features associated with the multi-view features extracted in Section~\ref{sec: mv} (i.e., $\boldsymbol{Z}=\boldsymbol{Z}_\text{mv}$). Then we employ the k-nearest neighbors (KNN) algorithm to determine the hyperedges. Specifically, we first calculate the feature similarity between any two flows, which is defined by the Euclidean distance between their feature vectors. Then for each flow, we assign the $K$ most similar flows and itself to a flow group, forming a hyperedge. 
The weight of each hyperedge is initialized to 1 (i.e., $\boldsymbol{M} = \boldsymbol{I}$).
This hyperedge construction strategy facilitates the aggregation of similar flow instances in the feature space, thereby enhancing the local structural representation of minority-class instances and effectively alleviating the data imbalance issue. In contrast to the generic hyperedge construction used in HyperGCN~\cite{yadati2019hypergcn}, \model incorporates prior knowledge of traffic behavior into the hyperedge generation process. By integrating temporal features with a class-aware aggregation strategy, \model achieves superior detection performance on imbalanced datasets.

The hypergraph construction in \model is specifically tailored for network traffic analysis, grounded in the assumption of intra-class similarity, \ie flow instances of the same class exhibit higher similarity in multi-view feature space, whereas inter-class similarity remains low. Unlike existing methods such as HyperGCN that rely on static features, \model leverages temporal multi-view features, including temporal patterns, interaction behaviors, and payload characteristics. These features are derived through time-series analysis and real-time session processing, enabling the capture of temporal traffic behaviors. This temporal modeling strategy provides a significant advantage over the static feature modeling.

\subsection{Traffic Hypergraph Learning}
We formulate the problem of traffic detection as a node classification task in a hypergraph, wherein we learn the mapping between flow node features and labels by constructing a hypergraph encoding model. Furthermore, to address the challenges posed by imbalanced class distribution and label scarcity, we introduce a contrastive self-supervision mechanism and design dual-contrastive constraints as proxy tasks to optimize the hypergraph encoding model.

\subsubsection{Hypergraph Encoder Architecture}
The constructed traffic hypergraph explicitly represents flow groups, revealing the clustering relationships between flows. This helps in uncovering the complex correlations of network activities. For the constructed traffic hypergraph, we further adopt a hypergraph convolutional network (HyperGCN)~\cite{yadati2019hypergcn} as encoder to learn the latent features of the flows and flow groups. The HyperGCN achieves message passing through hyperedges, which can be divided into two processes: 1) \textbf{Feature propagation from flows to flow groups ($\mathcal{V}\rightarrow\mathcal{E}$):} Each flow propagates its features to the flow groups it belongs to, and each flow group aggregates the features of its constituent flows; 2) \textbf{Feature propagation from flow groups to flows ($\mathcal{E}\rightarrow\mathcal{V}$):} After aggregating the features of the contained flows, the flow group propagates the aggregated features back to all its associated flows, completing the feature update. The processes of feature update during message passing can be defined as follows:
\begin{equation}
    \begin{aligned}  
    &\mathcal{V}\rightarrow\mathcal{E}: \quad \boldsymbol{E}^{(l)}=\sigma\left(\boldsymbol{D}_e^{-1}\boldsymbol{H}^T\boldsymbol{V}^{(l-1)}\boldsymbol{W}_{e}^{(l)}+\boldsymbol{b}_{e}^{(l)}\right)\\
    &\mathcal{E}\rightarrow\mathcal{V}: \quad
    \boldsymbol{V}^{(l)}=\sigma\left(\boldsymbol{D}_v^{-1}\boldsymbol{H}\boldsymbol{M}\boldsymbol{E}^{(l)}\boldsymbol{W}_{v}^{(l)}+\boldsymbol{b}_{v}^{(l)}\right)
    \end{aligned}
\end{equation}
where $\boldsymbol{V}^{(0)}=\boldsymbol{Z}$, $\boldsymbol{V}^{(l)}$ and $\boldsymbol{E}^{(l)}$ denote the learned representations of the nodes (flows) and hyperedges (flow groups) during the $l$-th message passing respectively, $\boldsymbol{W}_{e}$ and $\boldsymbol{W}_{n}$ are the weight matrices, and $\boldsymbol{b}_{e}^{(l)}$ and $\boldsymbol{b}_{v}^{(l)}$ are trainable biases.

\subsubsection{Hypergraph Data Augmentation}


Contrastive learning relies on data augmentation~\cite{zhou2022data,Mevolve,zhou2020data-cikm} to generate contrastive views. Here we design three data augmentation strategies: node feature masking, hypergraph weight perturbation, and flow-group membership masking. 
By introducing random perturbations, these strategies enrich data diversity, thereby alleviating data imbalance and label scarcity issues, and improving model performance across diverse traffic scenarios.
\begin{itemize}[leftmargin=15pt]
    \item \textbf{Node Feature Masking (NF).} We first generate a random binary mask of size $|\mathcal{V}|$, with its elements sampled from a Bernoulli distribution $\mathcal{B}(1-p_n)$. This mask is used to determine which nodes' features should be masked with an all-zero vector.
    \item \textbf{Hyperedge Weight Perturbation (EW).} We first generate a random binary mask of size $|\mathcal{E}|$, with its elements sampled from a Bernoulli distribution $\mathcal{B}(1-p_w)$, where $p_w$ is the perturbation probability. We then use this mask to determine whether the weight of a hyperedge should be perturbed, then apply Gaussian noise to replace the perturbed weights.
    \item \textbf{Flow-group Membership Masking (ED).} We first generate a mask of the same size as the incidence matrix, with its elements sampled from a Bernoulli distribution $\mathcal{B}(1-p_m)$. This mask is then used to determine whether the membership relationship between a flow and its associated group should be masked.
\end{itemize}
For the flow hypergraph, we generate two augmented views for contrastive learning by applying one or more data augmentation strategies. By combining different augmentation strategies, various view pairs can be obtained, and their effects will be discussed in the experimental analysis.

\subsubsection{Hypergraph Contrastive Learning}
With the hypergraph encoder, we can effectively propagate and update the multi-view features of flows within flow hypergraph, thereby enabling the incorporation of higher-order correlations into the updated flow features. However, traffic data in different scenarios inherently exhibit class imbalances, which may lead to fairness issues during model training. For instance, in scenarios involving DDos attacks, the ratio of normal to malicious traffic is extremely skewed, causing the model to be biased toward normal traffic during training, which in turn reduces its ability to detect abnormal traffic. Meanwhile, the labeling of traffic data is costly and challenging, leading to label scarcity, which further causes the model to fall into overfitting during training.

To facilitate effective learning of generalized flow features in scenarios with limited data labels and imbalanced distributions, we further introduce the contrastive self-supervised mechanism and design dual-contrastive proxy tasks: flow-flow (node-node) contrast and group-group (hyperedge-hyperedge) contrast. We employ the dual-contrastive constraints as regularization to jointly optimize the hypergraph encoder.

In the hypergraph contrastive learning setting, the flow hypergraph will undergo two data augmentation transformations, $T_1$ and $T_2$, to generate two contrastive views $\hat{\mathcal{H}}_{(1)}$ and $\hat{\mathcal{H}}_{(2)}$, respectively. These two contrastive views are fed into the hypergraph encoder, generating the corresponding representations of flows and flow groups: $(\hat{\boldsymbol{V}}_{(1)}^{(l)}, \hat{\boldsymbol{E}}_{(1)}^{(l)})$ and $(\hat{\boldsymbol{V}}_{(2)}^{(l)}, \hat{\boldsymbol{E}}_{(2)}^{(l)})$. To enable more efficient contrastive optimization, we then use a projection head, consisting of a two-layer Multi-Layer Perceptron (MLP) and an ELU activation function, to further map these two hypergraph representations into a low-dimensional space, yielding projected representations $(\hat{\boldsymbol{V}}_{(1)}, \hat{\boldsymbol{E}}_{(1)})$ and $(\hat{\boldsymbol{V}}_{(2)}, \hat{\boldsymbol{E}}_{(2)})$ for subsequent contrast.

The \textbf{flow-to-flow (node-to-node) contrast} encourages the model to produce similar representations for the same flow instance under different views, while increasing the representational distance from other flow instances. This intra-class similarity assumption significantly enhances the model's ability to represent minority-class samples, thereby alleviating issues related to data imbalance and label scarcity.
Specifically, for each flow $v_i$, we treat its representation in view 1 as the anchor, its representation in view 2 as the positive sample, and the representations of other flows in view 2 as negative samples. The constraint for each flow can be formulated as:
\begin{equation}
    \ell_n(\boldsymbol{\hat{V}}_{1,i},\ \boldsymbol{\hat{V}}_{2,i})=-\log\frac{e^{\cos(\boldsymbol{\hat{V}}_{1,i},\ \boldsymbol{\hat{V}}_{2,i})/\tau_n}}{\sum_{t=1}^{|\mathcal{V}|}e^{\cos(\boldsymbol{\hat{V}}_{1,i},\ \boldsymbol{\hat{V}}_{2,t})/\tau_n}}
\end{equation}
where $\cos(\cdot)$ is the cosine similarity function, and $\tau_n$ is the temperature parameter. In this contrast, we can swap the anchor and the positive sample, treating the representation in view 2 as the anchor, the corresponding representation in view 1 as the positive sample, and the representations of other flows in view 1 as the negative samples, resulting in a symmetric contrast. The final complete flow-flow contrastive constraint can be achieved by minimizing the following loss function:
\begin{equation}
    \mathcal{L}_n=\frac{1}{2|\mathcal{V}|}\sum_{i=1}^{|\mathcal{V}|}\left\{\ell_n(\boldsymbol{\hat{V}}_{1,i},\ \boldsymbol{\hat{V}}_{2,i})+\ell_n(\boldsymbol{\hat{V}}_{2,i},\ \boldsymbol{\hat{V}}_{1,i})\right\}
\end{equation}


The \textbf{group-to-group (hyperedge-to-hyperedge) contrast}  optimizes the hypergraph encoder by modeling the similarity among groups of similar flows and the dissimilarity across different flow groups. This facilitates the learning of high-order relational patterns at a macro level, reduces the noise introduced by KNN-based hypergraph construction, and further alleviates the challenges posed by imbalanced data and label scarcity.
Specifically, for each flow group $e_j$, we treat its representation in view 1 as the anchor, its representation in view 2 as the positive sample, and the representations of other flow groups in view 2 as negative samples. The constraint for each flow group is formulated as:
\begin{equation}
    \ell_g(\boldsymbol{\Hat{E}}_{1,j},\ \boldsymbol{\Hat{E}}_{2,j})=-\log\frac{e^{\cos(\boldsymbol{\Hat{E}}_{1,j},\ \boldsymbol{\Hat{E}}_{2,j})/\tau_g}}{\sum_{t=1}^{|\mathcal{E}|}e^{\cos(\boldsymbol{\Hat{E}}_{1,j},\ \boldsymbol{\Hat{E}}_{2,t})/\tau_g}}
\end{equation}
where $\tau_g$ is the temperature parameter. Similarly, in this contrast mode, we also adopt a symmetric form. The final complete group-to-group contrastive constraint can be achieved by minimizing the following loss function:
\begin{equation}
    \mathcal{L}_{g}=\frac{1}{2|\mathcal{E}|}\sum_{j=1}^{|\mathcal{E}|}\left\{\ell_{g}(\boldsymbol{\Hat{E}}_{1,j},\ \boldsymbol{\Hat{E}}_{2,j})+\ell_{g}(\boldsymbol{\Hat{E}}_{2,j},\ \boldsymbol{\Hat{E}}_{1,j})\right\}
\end{equation}

The aforementioned dual-contrastive optimizations can effectively improve the encoder's ability to distinguish flow features, enhancing its robustness in noisy network environments, as well as its generalization capability in scenarios with limited labels and imbalanced data.

\subsection{Model Training}

The process of framework training proceeds as follows:
1) Apply data augmentation strategies to the original flow hypergraph $\mathcal{H}$ to generate two augmented views $\hat{\mathcal{H}}_{(1)}$ and $\hat{\mathcal{H}}_{(2)}$;
2) Input these three graphs ($\mathcal{H}$, $\hat{\mathcal{H}}_{(1)}$, $\hat{\mathcal{H}}_{(2)}$) into the encoder for representation learning, obtaining the embeddings $(\boldsymbol{V}^{(l)}, \boldsymbol{E}^{(l)})$, $(\hat{\boldsymbol{V}}^{(l)}_{(1)}, \hat{\boldsymbol{E}}^{(l)}_{(1)})$, $(\hat{\boldsymbol{V}}^{(l)}_{(2)}, \hat{\boldsymbol{E}}^{(l)}_{(2)})$;
3) For the flow representations $\boldsymbol{V}^{(l)}$ from the original graph, we use a prediction head (two-layer MLP with softmax activation function) to obtain the predicted distribution $\tilde{\boldsymbol{y}}\in\mathbb{R}^{C}$ of the flows, and compute the cross-entropy loss:
\begin{equation}
    \mathcal{L}_\text{pred}=-\frac{1}{N}\sum_{i=1}^N\sum_{j=1}^C 
 \boldsymbol{y}_{i,j}\log(\tilde{\boldsymbol{y}}_{i,j})
\end{equation} 
where $C$ is the number of classes, $\boldsymbol{y}$ is the one-hot encoded true label;
4) For the representations from the augmented graphs, we use the prediction head to map them into a lower-dimensional space for contrastive learning and compute the dual-contrastive loss;
5) The dual-contrastive constraint is treated as an auxiliary task, and it is jointly trained with the traffic detection (classification) task. The entire framework is optimized by minimizing the following loss function:
\begin{equation}\label{eq: loss}
    \mathcal{L}=
    \mathcal{L}_\text{pred}+
    \omega_n\cdot\mathcal{L}_n+
    \omega_g\cdot\mathcal{L}_g
\end{equation}
where $\omega_n$ and $\omega_g$ represent the weight parameters of $\mathcal{L}_n$ and $\mathcal{L}_g$ respectively.

\subsection{{Online Traffic Detection}}


The core encoder of \model leverages inductive learning by constructing three distinct and independent hypergraphs as training, validation, and testing sets, respectively, ensuring robust generalization to unseen data. This design confers significant potential for \model's deployment in online traffic detection scenarios. Users can capture network traffic within predefined time windows to generate hypergraph snapshots, which \model can then analyze for real-time detection. Moreover, \model's multi-view feature extraction module is decoupled from its detection module, enabling pre-trained parallel processing of hypergraph snapshots from different time windows, thus substantially improving the efficiency of online detection.

In practical network environments, \model's design effectively addresses multiple challenges. Unlike traditional graph-based methods that struggle with dynamic maintenance due to data accumulation, \model's inductive approach significantly alleviates storage pressure through dynamic slicing. Furthermore, its multi-view feature extraction module comprehensively captures the complexities inherent in mixed traffic, making it adaptable to diverse data scenarios. To address real-world noise and complexity, \model incorporates a dual contrastive learning mechanism, enhancing feature representations of individual traffic flows and their corresponding groups. This approach markedly boosts the robustness and detection performance of the model.

\begin{algorithm}[!h]
\caption{FlowID Algorithm}
\label{alg:FlowID}
\renewcommand{\algorithmicrequire}{\textbf{Input:}}
\renewcommand{\algorithmicensure}{\textbf{Output:}}
\begin{algorithmic}[1]
    \REQUIRE Traffic hypergraph $\mathcal{H}=(\mathcal{V},\mathcal{E},\boldsymbol{M},\boldsymbol{H},\boldsymbol{Z})$  
    \ENSURE Optimized embeddings $\boldsymbol{V}^{(l)}$, traffic predictions $\hat{y}$  
    \STATE Apply data augmentation to generate two augmented views $\hat{\mathcal{H}}_{(1)}$ and $\hat{\mathcal{H}}_{(2)}$ from $\mathcal{H}$;
    \STATE Input ($\mathcal{H}$, $\hat{\mathcal{H}}_{(1)}$, $\hat{\mathcal{H}}_{(2)}$) into the encoder to obtain embeddings (${\boldsymbol{V}}^{(l)}, \hat{\boldsymbol{V}}^{(l)}_{(1)}, \hat{\boldsymbol{V}}^{(l)}_{(2)})$;
    \FOR{each flow $v_i$ in $\mathcal{H}$}
        \STATE Use the prediction head on ${\boldsymbol{V}}^{(l)}$ to predict labels $\tilde{\boldsymbol{y}}_i$;
        \STATE Compute the cross-entropy loss $\mathcal{L}_{\text{pred}}$;
    \ENDFOR
    \FOR{each augmented view $(\hat{\boldsymbol{V}}^{(l)}_{(1)}, \hat{\boldsymbol{V}}^{(l)}_{(2)})$}
        \STATE Project embeddings into lower-dimensional space;
        \STATE Compute the contrastive loss $\mathcal{L}_{n}$ and $\mathcal{L}_{g}$;
    \ENDFOR
    \STATE Optimize encoder by minimizing the loss $\mathcal{L}$ in Eq.~(\ref{eq: loss});
    \RETURN Optimized embeddings ${\boldsymbol{V}}^{(l)}$, predicted labels $\hat{y}$.
\end{algorithmic}
\end{algorithm}

\section{Experiment}\label{sec: experiment}

\subsection{Dataset}
We evaluate the performance of \model ~on five widely used datasets: CIC-IOMT2024~\cite{dadkhah2024ciciomt2024}, UNSW-NB15~\cite{moustafa2015unsw}, Darknet2020~\cite{habibi2020didarknet}, ISCX-VPN2016~\cite{draper2016characterization}, and USTC-TFC2016~\cite{wang2017malware}. The statistical information of datasets is shown in Table~\ref{tab: dataset}.

CIC-IOMT2024~\cite{dadkhah2024ciciomt2024} was released by the Canadian Institute for Cybersecurity and focuses on enhancing security in the realm of Internet of Medical Things (IoMT). It comprises data from 40 devices, including 25 real devices and 15 simulated ones, encompassing a total of 18 attacks across 5 distinct categories. It encompasses a wide array of healthcare communication protocols such as Wi-Fi, MQTT, and Bluetooth.

\begin{table}[!htb] 
    \renewcommand\arraystretch{1.3} 
    \centering                 
    \caption{{Statistical information of traffic datasets.}}
    \resizebox{\linewidth}{!}{
    \begin{tabular}{cccccc} 
\hline
Datasets     & \begin{tabular}[c]{@{}c@{}}Num. Flows \\$|\mathcal{V}|$\end{tabular} & \begin{tabular}[c]{@{}c@{}}Num. groups\\$|\mathcal{E}|$\end{tabular} &Membership  &Max.$D_v$  & Class  \\ 
\hline
CIC-IOMT2024 &32,502 &32,502 &97,506  &458  &6        \\
UNSW-NB15    &24,665 &24,665 &73,995  &1229  &5        \\
Darknet2020  &35,533 &35,533 &106,599  &1804  &7        \\
ISCX-VPN2016 &18,232 &18,232 &54,696  &1105  &6        \\
USTC-TFC2016  &65,602 &65,602 &196,805  &45  &11        \\
\hline
\end{tabular}}
\label{tab: dataset} 
\end{table}

UNSW-NB15~\cite{moustafa2015unsw} was constructed by the Cyber Range Labs at the University of New South Wales using the IXIA PerfectStorm tool in 2015, comprises raw network packets that encompass both genuine network normal activity and attack behavior. It encompasses 9 distinct attack types. The anomalous behavior of UNSW-NB15 dataset is more novel and balanced, which is suitable for network anomalous traffic detection research.

Darknet2020~\cite{habibi2020didarknet} integrates two publicly available encrypted datasets (ISCX-VPN2016 and ISCX-Tor2016) to form a comprehensive darknet traffic dataset. The encrypted traffic encompasses two types of encrypted communication methods, namely Tor and VPN, respectively, and encompasses six diverse types of encrypted application traffic. 

ISCX-VPN2016~\cite{draper2016characterization} comprises encrypted communication traffic transmitted via virtual private network (VPN) tunnels, encompassing five distinct applications such as VoIP and email. VPNs are frequently employed for circumventing censorship or obfuscating hidden locations through protocols, commonly facilitating access to blocked websites or services.

USTC-TFC2016~\cite{wang2017malware} was released by a research team affiliated with the University of Science and Technology of China in 2017. It comprises encrypted network traffic encompassing both malicious and benign applications. Specifically, the malicious traffic encompasses ten distinct viruses or Trojans, including Cridex, Geodo, and Htbot.

\subsection{Comparison Methods}\label{sec: baseline}
We compare \model with nine network traffic detection methods from two categories: sequence-based methods (CNN~\cite{zhang2019framework}, LSTM~\cite{hwang2019lstm}, App-Net~\cite{wang2020app}) and graph-based methods (TCGNN~\cite{hu2023tcgnn}, HGNN~\cite{feng2019hypergraph}, HGNN+~\cite{gao2022hgnn+}, GraphDApp~\cite{shen2021accurate}, GraphDDoS~\cite{li2022graphddos}, TFE-GNN~\cite{TFE-GNN}).

CNN~\cite{zhang2019framework}: This method utilizes a 1-D convolutional neural network to extract key information from packet payload sequences, thereby achieving traffic classification.

LSTM~\cite{hwang2019lstm}: This method uses LSTM to extract temporal patterns from the packet length sequence of the traffic, thereby achieving traffic classification.

App-Net~\cite{wang2020app}: This method utilizes a hybrid neural network that combines Bi-LSTM and CNN in a parallel structure. This design allows it to learn both packet length sequences and initial data packet signatures in TLS flows, extracting joint features for more accurate encrypted traffic classification.

TCGNN~\cite{hu2023tcgnn}: This method first transforms network packets into graphs, where bytes serve as nodes and their associations are modeled as edges, and then employs a two-layer GCN to captures packet-level features, achieving highly accurate classification of network traffic.

HGNN~\cite{feng2019hypergraph}: This method uses a hypergraph neural network (HGNN) framework that incorporates hypergraph structures to model complex and high-order data correlations. By designing a hyperedge convolution operation, it efficiently captures relationships in multi-modal data, enabling more effective representation learning compared to traditional GNNs. 

HGNN+~\cite{gao2022hgnn+}: This method introduces an adaptive fusion strategy to combine different hyperedge groups, which enhances the representation of multi-type data. By employing a new hypergraph convolution scheme in the spatial domain, it efficiently learns a unified representation for various tasks.

\begin{table*}[!htb] 
    \renewcommand\arraystretch{1.3} 
    \centering                   
    \caption{Performance comparison of \model with baselines across different datasets. The best results are highlighted in bold, and the second-best results are underlined. ``Average ranking'' represents the average ranking of the metrics achieved by a method across all datasets.} 
    \resizebox{\linewidth}{!}{
    \begin{tabular}{c|c|cccccccccc} 
\hline
Datasets                                                               & Method & CNN                & LSTM       & HGNN       & HGNN+               & TCGCN      & App-Net            & GraphDDoS  & GraphDApp  & TFE-GNN            & \model                 \\ 
\hline
\multirow{4}{*}{CIC-IOMT2024}                                          & ACC    & 93.78±0.27         & 94.05±0.13 & 94.83±0.05 & \uline{94.89±0.03}  & 90.63±0.29 & 93.79±0.40         & 90.35±0.06 & 81.23±0.19 & 93.09±0.14         & \textbf{95.27±0.01}  \\
                                                                       & M-Pr   & 88.49±2.00         & 88.72±0.44 & 93.71±0.30 & \textbf{93.97±0.19} & 82.15±1.50 & 88.37±2.61         & 70.45±9.61 & 51.48±0.72 & 76.82±4.56         & \uline{93.75±0.46}   \\
                                                                       & M-Re   & 84.45±0.43         & 84.66±0.60 & 85.06±0.29 & \uline{86.08±0.09}  & 68.13±1.55 & 83.69±1.27         & 60.75±1.29 & 49.84±0.50 & 56.70±1.10         & \textbf{88.89±0.12}  \\
                                                                       & M-F1   & 86.03±0.92         & 86.43±0.56 & 88.39±0.16 & \uline{89.17±0.11}  & 72.22±1.47 & 85.61±1.75         & 62.14±2.32 & 49.64±0.57 & 59.70±1.31         & \textbf{90.36±0.10}  \\ 
\hline
\multirow{4}{*}{UNSW-NB15}                                             & ACC    & 94.06±0.08         & 94.35±0.17 & 94.39±0.07 & \uline{94.41±0.03}  & 91.94±0.28 & 94.20±0.12         & 92.32±0.22 & 79.36±0.20 & 92.18±0.40         & \textbf{94.49±0.03}  \\
                                                                       & M-Pr   & 84.20±1.15         & 87.48±0.71 & 87.96±1.46 & \uline{89.39±0.13}  & 72.38±4.41 & 85.95±1.21         & 83.33±2.37 & 49.11±1.71 & 86.20±2.78         & \textbf{89.78±0.21}  \\
                                                                       & M-Re   & 83.04±2.37         & 82.78±1.16 & 82.38±0.31 & 83.58±0.17          & 66.67±2.44 & \uline{83.72±1.41} & 68.75±3.81 & 44.95±0.32 & 66.04±3.03         & \textbf{84.30±0.17}  \\
                                                                       & M-F1   & 83.15±0.89         & 84.54±0.46 & 84.48±0.39 & \uline{85.51±0.09}  & 67.78±2.29 & 84.54±1.05         & 71.73±1.51 & 43.95±0.56 & 66.13±3.52         & \textbf{86.05±0.09}  \\ 
\hline
\multirow{4}{*}{Darknet2020}                                           & ACC    & 93.85±0.14         & 94.04±0.02 & 93.89±0.15 & \uline{94.29±0.05}  & 89.75±0.32 & 94.11±0.39         & 91.22±0.23 & 79.29±0.20 & 85.46±0.74         & \textbf{94.51±0.06}  \\
                                                                       & M-Pr   & 81.81±0.24         & 80.42±1.55 & 84.55±0.94 & \uline{84.81±0.78}  & 76.41±4.44 & 82.22±0.77         & 71.29±0.52 & 42.24±0.18 & 79.51±1.02         & \textbf{86.41±0.50}  \\
                                                                       & M-Re   & 77.28±1.34         & 78.25±0.09 & 77.55±1.10 & \uline{80.57±0.25}  & 53.16±2.03 & 78.47±1.10         & 67.32±0.88 & 37.58±0.40 & 77.84±2.23         & \textbf{82.62±0.48}  \\
                                                                       & M-F1   & 79.06±1.00         & 79.11±0.91 & 79.93±0.53 & \uline{81.00±0.15}  & 57.10±1.04 & 79.75±0.96         & 68.66±0.80 & 38.12±0.38 & 78.14±0.56         & \textbf{82.44±0.22}  \\ 
\hline
\multirow{4}{*}{ISCX-VPN2016}                                          & ACC    & 84.06±0.09         & 83.06±0.59 & 86.00±0.36 & 86.75±0.03          & 75.64±0.25 & 83.77±0.07         & 75.27±1.28 & 76.28±0.35 & \uline{87.50±0.81} & \textbf{87.97±0.07}  \\
                                                                       & M-Pr   & 80.95±0.52         & 79.11±2.26 & 89.08±0.84 & \textbf{93.07±0.55} & 65.20±2.19 & 78.72±1.76         & 71.90±0.90 & 52.40±0.68 & 87.87±1.03         & \uline{91.37±0.23}   \\
                                                                       & M-Re   & 74.63±0.41         & 78.40±2.74 & 75.59±0.99 & 79.28±0.36          & 51.49±3.20 & 75.51±1.53         & 58.92±3.19 & 45.28±0.92 & \uline{85.23±0.95} & \textbf{90.22±0.15}  \\
                                                                       & M-F1   & 76.67±0.79         & 78.04±0.02 & 79.63±0.65 & 82.01±0.23          & 55.83±2.55 & 75.94±0.82         & 59.36±1.21 & 46.70±0.90 & \uline{86.32±1.11} & \textbf{88.33±0.12}  \\ 
\hline
\multirow{4}{*}{USTC-TFC2016}                                          & ACC    & \uline{97.60±0.08} & 97.17±0.21 & 96.07±0.08 & 97.44±0.04          & 95.78±0.19 & 97.36±0.20         & 93.28±0.37 & 48.95±0.19 & 90.24±5.70         & \textbf{97.67±0.04}  \\
                                                                       & M-Pr   & 96.72±0.10         & 95.66±0.16 & 91.02±4.12 & \uline{96.95±0.07}  & 88.97±0.33 & 96.42±0.25         & 88.85±0.54 & 26.88±0.16 & 77.91±1.32         & \textbf{96.98±0.10}  \\
                                                                       & M-Re   & \uline{96.82±0.28} & 95.95±0.47 & 83.72±0.98 & 96.19±0.22          & 85.75±0.66 & 96.41±0.36         & 87.50±1.46 & 37.49±0.30 & 75.34±2.78         & \textbf{96.99±0.08}  \\
                                                                       & M-F1   & \uline{96.76±0.14} & 95.78±0.19 & 85.42±1.07 & 96.47±0.11          & 86.99±0.56 & 96.40±0.29         & 87.13±0.58 & 30.69±0.17 & 72.14±0.80         & \textbf{96.90±0.06}  \\ 
\hline
\multirow{4}{*}{\begin{tabular}[c]{@{}c@{}}Average\\Rank\end{tabular}} & ACC    & 5                  & 4.8        & 4.2        & 2.4                 & 8.2        & 4.6                & 8.2        & 9.6        & 7                  & 1                    \\
                                                                       & M-Pr   & 5                  & 5          & 3.6        & 1.6                 & 8          & 5.4                & 8.4        & 10         & 6.6                & 1.4                  \\
                                                                       & M-Re   & 5                  & 4.4        & 5.6        & 2.8                 & 8          & 4                  & 7.4        & 10         & 6.8                & 1                    \\
                                                                       & M-F1   & 5                  & 4.4        & 4.6        & 2.4                 & 8          & 4.8                & 7.4        & 10         & 7.2                & 1                    \\
\hline
\end{tabular}}
\label{tab: result} 
\end{table*}

GraphDDoS~\cite{li2022graphddos}: This method leverages a GNN to detect DDoS attacks by constructing endpoint traffic graphs that capture the relationships between packets and flows. By designing a GNN classifier that analyzes both packet-level interactions and flow-level patterns, it effectively identifies abnormal traffic behaviors characteristic of DDoS attacks. 

GraphDApp~\cite{shen2021accurate}: This method constructs each flow as a graph, where nodes represent packets and edges represent interactions between packets. It models encrypted traffic detection as a graph classification task.

TFE-GNN~\cite{TFE-GNN}: This method constructs a graph by mining the correlation between bytes, where each byte corresponds to a node in the graph. The PMI value is used to determine the connection between nodes, and a graph neural network and a cross-gated fusion mechanism are used to learn the feature representation of encrypted traffic.

\subsection{Evaluation Metrics}\label{sec: metrics}
Considering the class imbalance in traffic datasets, we use the macro-averaged versions of Accuracy, Precision, Recall, and F1-score as evaluation metrics to comprehensively measure the performance of different methods. The reason for this is that the macro-averaged metrics first calculate the metric for each class independently and then average the results across all classes, ensuring that the performance is not dominated by the majority classes. Therefore, it better reflects the model's performance on minority classes. The calculation of the macro-averaged metrics is defined as follows:
\begin{equation} 
    \text{Macro-averaged Metric}=\frac1C\sum_{i=1}^C \text{ Metric}_i
\end{equation}
where $\text{Metric}_i$ is the metric of class $i$.

\subsection{Experimental Setting}\label{sec: Hyperparameters}
\model is implemented using PyTorch and Pytorch-Geometric library. 
During feature extraction, we employ a two-layer GCN, a single layer LSTM, and a two-layer CNN, with an output dimension of 512. For CNN, we set the size of the 1D convolutional kernel to 25, stride to 1, and padding to 12. During hypergraph learning, we utilize two-layers HyperGCN, with both the hidden and projection dimensions set to 128, and Dropout rate to 0.2. Additionally, we use the ReLU activation function for non-linear transformations. During framework training, we use the Adam optimizer with a learning rate of 0.002 and a weight decay rate of 1e-3. In the experimental parameters, we set the number of neighbors $K$ in the hyperedge to 3, the number of intercepted packets m to 40, and the number of payload bytes to 16. All experiments will be repeated 10 times, and the metrics along with the corresponding standard deviations will be reported.

\begin{table}[!htb] 
    \renewcommand\arraystretch{1.3} 
    \centering                  
    \caption{The detection results of \model for different types of traffic.} 
    \resizebox{\linewidth}{!}{
    \begin{tabular}{cccc} 
\hline
CIC-IOMT2024 & Precision   & Recall      & F1-score     \\ 
\hline
Benign       & 95.30$\pm$0.34  & 99.83$\pm$0.04  & 97.08$\pm$0.03   \\
ARP          & 98.75$\pm$3.31  & 94.13$\pm$0.21  & 87.53$\pm$0.94   \\
DDoS         & 93.02$\pm$0.02  & 98.91$\pm$0.08  & 95.64$\pm$0.01   \\
DoS          & 100.00$\pm$0.00 & 91.99$\pm$0.01  & 95.53$\pm$0.00   \\
MQTT         & 100.00$\pm$0.00 & 88.11$\pm$0.71  & 92.61$\pm$0.17   \\
Recon        & 96.56$\pm$1.82  & 63.65$\pm$4.12  & 69.46$\pm$0.45   \\ 
\hline\hline
UNSW-NB15    & Precision   & Recall      & F1-score     \\ 
\hline
Benign       & 96.00$\pm$2.80  & 93.23$\pm$3.84  & 90.38$\pm$0.73   \\
DoS          & 60.87$\pm$47.23 & 58.42$\pm$5.15  & 66.25$\pm$1.59   \\
Exploits     & 97.56$\pm$1.65  & 97.07$\pm$0.83  & 94.85$\pm$0.25   \\
Generic      & 92.68$\pm$0.21  & 99.80$\pm$0.11  & 95.78$\pm$0.04   \\
ShellCode    & 100.00$\pm$0.00 & 91.21$\pm$0.31  & 95.48$\pm$0.16   \\ 
\hline\hline
Darknet2020  & Precision   & Recall      & F1-score     \\ 
\hline
Browsing     & 96.42$\pm$0.60  & 99.58$\pm$0.30  & 95.44$\pm$0.04   \\
Chat         & 53.63$\pm$42.05 & 63.23$\pm$4.19  & 47.20$\pm$1.83   \\
Email        & 42.02$\pm$35.27 & 93.75$\pm$1.29  & 89.36$\pm$0.31   \\
File         & 99.68$\pm$0.33  & 94.98$\pm$1.61  & 93.31$\pm$0.14   \\
P2P          & 99.73$\pm$0.08  & 99.20$\pm$0.33  & 99.00$\pm$0.03   \\
Streaming    & 86.79$\pm$3.75  & 86.91$\pm$2.95  & 73.92$\pm$0.57   \\
VOIP         & 85.88$\pm$3.09  & 87.57$\pm$0.81  & 78.31$\pm$0.60   \\ 
\hline\hline
ISCX-VPN2016 & Precision   & Recall      & F1-score     \\ 
\hline
Chat         & 87.48$\pm$049   & 99.11$\pm$0.19  & 79.26$\pm$0.04   \\
Email        & 100.00$\pm$0.00 & 55.11$\pm$9.06  & 63.86$\pm$5.98   \\
File         & 99.75$\pm$066   & 88.18$\pm$0.72  & 87.98$\pm$0.16   \\
P2P          & 100.00$\pm$0.00 & 93.49$\pm$1.55  & 92.97$\pm$0.71   \\
Streaming    & 98.62$\pm$2.41  & 98.38$\pm$0.20  & 95.10$\pm$0.25   \\
VOIP         & 98.81$\pm$0.41  & 99.87$\pm$0.16  & 91.02$\pm$0.04   \\ 
\hline\hline
USTC-TFC2016  & Precision   & Recall      & F1-score     \\ 
\hline
Benign       & 99.96$\pm$0.08  & 99.91$\pm$0.04  & 99.94$\pm$0.00   \\
Cridex       & 100.00$\pm$0.00 & 100.00$\pm$0.00 & 100.00$\pm$0.00  \\
Geodo        & 99.94$\pm$0.05  & 99.96$\pm$0.05  & 99.70$\pm$0.02   \\
Htbot        & 99.94$\pm$0.08  & 99.59$\pm$0.15  & 99.15$\pm$0.10   \\
Miuref       & 99.92$\pm$0.17  & 99.91$\pm$0.03  & 99.88$\pm$0.07   \\
Neris        & 96.41$\pm$1.85  & 95.43$\pm$1.38  & 91.59$\pm$0.09   \\
Nsis-ay      & 100.00$\pm$0.00 & 91.53$\pm$0.72  & 93.27$\pm$0.20   \\
Shifu        & 99.38$\pm$0.79  & 99.00$\pm$0.23  & 96.70$\pm$0.27   \\
Tinba        & 97.39$\pm$3.42  & 100.00$\pm$0.00 & 99.63$\pm$0.97   \\
Virut        & 59.44$\pm$38.72 & 95.35$\pm$1.00  & 87.38$\pm$0.16   \\
Zeus         & 75.00$\pm$43.30 & 98.76$\pm$0.24  & 97.96$\pm$0.27   \\
\hline
\end{tabular}}
\label{tab: multiclassification}  
\end{table}

\subsection{Traffic Detection Performance}\label{sec:results}

Table~\ref{tab: result} reports the traffic detection performance of ten methods across five datasets, from which we can draw the following conclusions:
1) \model significantly outperforms other methods in most cases, achieving the highest average performance ranking. This is reasonably attributed to its novel design: capturing multi-view features of flows as encoder inputs and emphasizing inter-traffic relationships through a hypergraph structure. It further employs dual-contrastive constraints to enhance discriminative representations among different traffic types, ultimately achieving state-of-the-art traffic detection performance; 
2) The performance rankings of all methods generally follow this order: hypergraph-based methods $>$ multi-view methods $>$ sequence-based methods $>$ graph-based methods. Hypergraph-based methods (\model, HGNN, HGNN+) also accept multi-view features as input and integrate higher-order correlations among flows, resulting in better traffic detection performance. Furthermore, multi-view methods (App-Net) combine both packet length sequence and payload sequence of traffic data, thus partially outperforming single-view methods in terms of overall performance.

\begin{figure}[!htb]
	\centering
	\includegraphics[width=\linewidth]{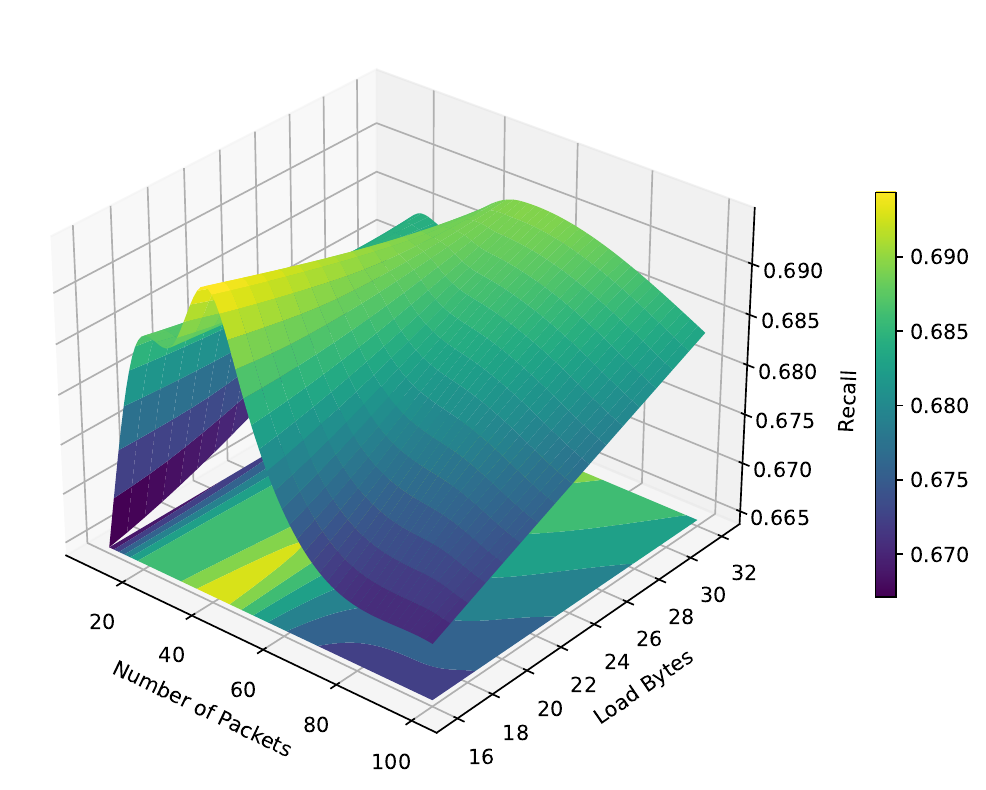}      
	\caption{The impact of the number of packets ($n$) and the number of payload bytes ($m$) used in the multi-view feature extraction process on the framework's performance.}
	\label{fig: 3D}
\end{figure}

Furthermore, Table~\ref{tab: multiclassification} reports the detection performance of \model for different types of traffic. Overall, \model achieves high precision and recall for most traffic categories, especially for normal traffic and certain types of attack traffic, where detection performance generally exceeds 95\%. For instance, on the USTC-TFC2016 dataset, \model exhibits outstanding detection performance for 9 out of 11 traffic types, with all three metrics surpassing 90\%. However, there is still room for improvement in identifying certain traffic categories. In the Recon attacks of the CIC-IOMT2024 dataset, the recall is only 63.65\%, and the F1-score is 69.46\%, indicating a risk of missed detections for this type of attack. For DoS attacks in the UNSW-NB15 dataset, the precision and recall are 60.87\% and 58.42\%, respectively, with the F1-score of 66.25\%, suggesting that further model optimization is needed to improve detection accuracy. Additionally, in the Email traffic of the Darknet2020 dataset, despite relatively high recall of 93.75\%, the precision is low (42.02\%), which may lead to a higher false positive rate. Overall, \model demonstrates strong capabilities in multi-category traffic detection, accurately detecting most traffic types. However, for categories with lower precision or recall, the framework requires further improvement to enhance the comprehensiveness and reliability of detection.



\begin{table*}[!htb] 
    \renewcommand\arraystretch{1.3} 
    \centering        
    \caption{Ablation analysis on CIC-IOMT2024, UNSW-NB15 and Darknet2020.}
    \resizebox{\textwidth}{!}{
\begin{tabular}{c|cccc|cccc|cccc} 
\hline
\multirow{2}{*}{Ablation} & \multicolumn{4}{c|}{CIC-IOMT2024}                                                     & \multicolumn{4}{c|}{UNSW-NB15}                                                                & \multicolumn{4}{c}{Darknet2020}                                                        \\ 
\cline{2-13}
                          & Accuracy            & Precision           & Recall        & F1            & Accuracy            & Precision     & Recall        & F1                    & Accuracy               & Precision     & Recall        & F1             \\ 
\hline   
\model w/o GNN            & 94.01±0.02          & 88.54±0.33          & 84.05±0.20          & 85.76±0.12          & 93.44±0.23          & 81.90±6.47          & 77.81±4.46          & 78.92±5.27                  & 94.18±0.40             & 81.22±1.92          & 74.48±2.90          & 76.00±2.15           \\
\model w/o CNN            & 93.43±0.08          & 89.17±0.43          & 83.05±0.23          & 85.37±0.17          & 94.22±0.09          & 86.39±1.54          & 81.78±0.87          & 82.25±0.37                  & 94.45±0.04             & 84.57±0.33          & 80.77±0.19          & \uline{82.30±0.11}   \\
\model w/o LSTM           & 93.80±0.05          & 90.02±0.99          & 84.54±0.44          & 85.73±0.16          & 93.74±0.22          & 80.73±1.73          & 78.95±2.64          & 79.46±2.28                  & 94.00±0.05             & 82.92±0.32          & 77.90±0.17          & 79.98±0.16           \\
\model Only GNN           & 88.56±0.05          & 61.78±3.11          & 53.41±0.07          & 54.21±0.29          & 91.86±0.05          & 68.82±2.69          & 65.36±0.55          & 65.77±0.51                  & 83.91±0.53             & 65.76±0.85          & 53.04±1.41          & 56.41±1.24           \\
\model Only LSTM          & 94.27±0.09          & 88.88±0.31          & 82.81±0.35          & 84.59±0.58          & 94.06±0.17          & 81.11±3.43          & 80.32±3.41          & 80.29±3.52                  & \textbf{94.85±0.35}    & 82.63±3.00          & 79.27±1.84          & 79.72±2.57           \\
\model Only CNN           & 93.67±0.04          & 88.90±0.82          & 81.26±0.51          & 83.78±0.31          & 93.72±0.11          & 83.40±0.81          & 75.01±0.90          & 77.00±0.96                  & 92.21±0.18             & 78.35±1.75          & 70.23±3.14          & 72.31±3.30           \\
\hline
\model Only MFE           & 94.21±0.15          & 89.38±1.09          & 85.34±1.26          & 86.19±1.20          & 94.12±0.24          & 84.77±1.67          & 83.86±1.71          & 84.01±1.34                  & \underline{94.80±0.35} & 84.61±0.18          & 81.03±0.51          & 81.34±1.06           \\
\model w/o Aug            & 95.04±0.07          & 93.17±0.51          & 87.21±0.47          & 88.99±0.45          & \uline{94.45±0.03}  & \uline{89.61±0.24}  & \uline{83.99±0.33}  & \uline{85.92±0.17}          & 94.43±0.06             & 84.58±1.25          & \uline{81.61±0.36}  & 81.58±0.39   \\
\model w/o CL             & \uline{95.05±0.05}  & \uline{93.73±0.34}  & \uline{87.31±0.31}  & \uline{89.12±0.27}  & 94.41±0.03          & 88.46±1.01          & 83.27±0.52          & 85.13±0.53                  & 94.30±0.11             & \uline{84.77±1.77}  & 81.23±0.70          & 80.75±0.82           \\
\hline   
\model                    & \textbf{95.27±0.01} & \textbf{93.75±0.46} & \textbf{88.89±0.12} & \textbf{90.36±0.10} & \textbf{94.49±0.03} & \textbf{89.78±0.21} & \textbf{84.30±0.17} & \textbf{86.05±0.09}         & 94.51±0.06             & \textbf{86.41±0.50} & \textbf{82.62±0.48} & \textbf{82.44±0.22}  \\
\hline
\end{tabular}}
\label{tab: Abaltion_mv1}
\end{table*}

\begin{table*}[!htb] 
    \renewcommand\arraystretch{1.3} 
    \centering   
    \caption{Ablation analysis on ISCX-VPN2016 and USTC-TFC2016.}
    \resizebox{0.71\textwidth}{!}{
    \begin{tabular}{c|cccc|cccc} 
\hline
\multirow{2}{*}{Ablation} & \multicolumn{4}{c|}{ISCX-VPN2016}                                                     & \multicolumn{4}{c}{USTC-TFC2016}                                                       \\ 
\cline{2-9}
                          & Accuracy            & Precision     & Recall        & F1            & Accuracy            & Precision     & Recall        & F1             \\ 
\hline
\model w/o GNN            & 84.54±0.40          & 87.28±0.81          & 75.50±1.42          & 79.33±0.79          & 96.68±0.39          & 94.72±3.45          & 90.54±3.33          & 91.52±3.49           \\
\model w/o CNN            & 83.32±0.12          & 83.17±0.50          & 74.86±0.64          & 76.28±0.63          & 97.28±0.03          & 96.05±0.08          & 96.40±0.16          & 96.13±0.07   \\
\model w/o LSTM           & 81.76±0.42          & 80.21±0.45          & 69.21±0.96          & 72.40±0.60          & 96.19±0.30          & 91.74±4.93          & 88.66±3.03          & 89.57±3.69           \\
\model Only GNN           & 70.65±1.73          & 69.00±2.29          & 50.09±0.91          & 53.26±0.94          & 88.74±0.23          & 83.92±2.19          & 74.48±1.67          & 77.53±2.08           \\
\model Only LSTM          & 80.09±0.09          & 83.18±0.22          & 70.61±0.95          & 74.33±0.77          & 96.90±0.13          & 95.22±0.74          & 94.69±0.64          & 94.70±0.58           \\
\model Only CNN           & 84.92±0.22          & 79.96±1.17          & 79.32±3.50          & 78.47±1.54          & 91.53±0.73          & 69.07±5.81          & 67.23±3.32          & 65.70±3.28           \\
\hline
\model Only MFE           & 85.06±0.22          & 85.92±1.58          & 78.24±2.50          & 78.64±2.35          & 97.59±0.20          & 96.49±0.56          & 96.83±0.30          & 96.68±0.35           \\
\model w/o Aug            & 87.93±0.08          & \uline{90.61±0.58}  & \uline{90.12±0.19}  & 88.23±0.26          & \uline{97.61±0.04}  & \uline{96.92±0.00}  & \uline{96.93±0.10}  & 96.81±0.07           \\
\model w/o CL             & \uline{87.98±0.05}  & 90.25±1.11          & 90.06±0.18          & \uline{88.35±0.13}  & 97.59±0.03          & 96.92±0.11          & 96.91±0.08          & \uline{96.84±0.08}   \\
\hline
\model                    & \textbf{88.04±0.05} & \textbf{91.44±0.46} & \textbf{90.25±0.12} & \textbf{88.36±0.10} & \textbf{97.67±0.04} & \textbf{96.98±0.10} & \textbf{96.99±0.08} & \textbf{96.90±0.06}  \\
\hline
\end{tabular}}
\label{tab: Abaltion_mv_2}
\end{table*}

\begin{figure*}[!htb]
	\centering
	\includegraphics[width=\textwidth]{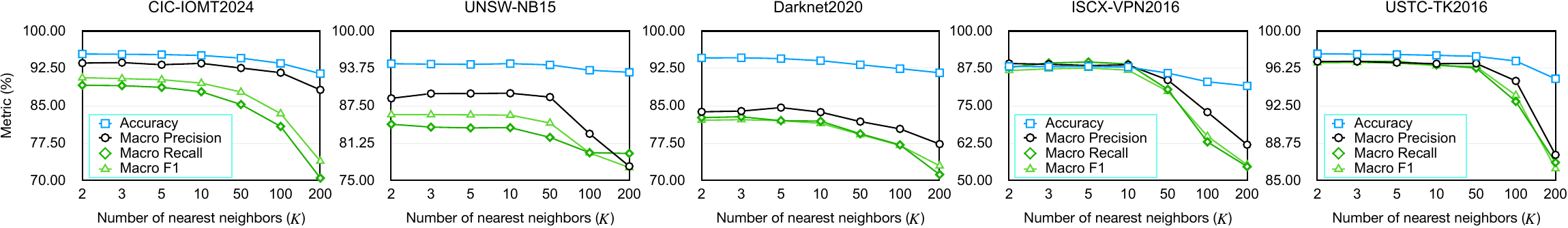}      
	\caption{The impact of the K-value setting during flow hypergraph construction on the framework's performance.}
	\label{fig: topk}
\end{figure*}

\subsection{Analysis on Multi-view Feature Extraction Module}

In this work, we use the first $n$ packets in a flow and the first $m$ bytes in the packet payload for feature extraction. Since the number of packets extracted from the flow and the number of payload bytes directly affect the training efficiency and performance of \model, we conduct an experimental analysis of these two parameters. According to \cite{bikmukhamedov2021multi}, excessively large values of $n$ and $m$ can lead to a significant decline in framework efficiency. Therefore, the parameter space of $n$ is set as \{10, 20, $\cdots$, 100\}, and the parameter space of $m$ is set as \{16, 18, $\cdots$, 32\}. By experimenting with various parameter combinations, we report the results in Fig.~\ref{fig: 3D}, illustrating how performance varies with different settings.

We can observe that as the number of packets $n$ increases from 10 to 40, the framework performance steadily improves, but beyond 40 packets, there is no further improvement in performance. This is because typical traffic flows contain no more than 40 packets, and when more packets are required, we fill the sequences with empty packets, resulting in meaningless data that negatively affects framework performance. 
As for the number of payload byte $m$, we find that with the same number of packets $n$, shorter payload byte information is more conducive to identifying different types of flows. This is because not all packets in flows carry payload data, and using too many bytes may introduce significant noise, degrading the detection ability.
Therefore, considering both framework efficiency and performance, we select this setting ($n = 40$, $m = 16$) as the final experimental parameters.


Furthermore, to investigate the impact of the multi-view feature extractor on the framework's performance, we design an ablation study across different views. Specifically, we sequentially remove single or multiple views from the framework and evaluate the model's performance changes on five datasets, as shown in Table ~\ref{tab: Abaltion_mv1} and ~\ref{tab: Abaltion_mv_2}. The experimental results demonstrate that removing any view leads to a performance degradation, while the multi-view approach significantly outperforms the single-view approach in most cases. 
This observation underscores the significance of deeply analyzing features such as packet size, transmission direction, payload patterns, and interaction behaviors in network traffic, all of which play a crucial role in enhancing the performance of detection frameworks. In particular, payload information is essential for capturing application-layer semantics and user behavior. Results show that removing the payload-related feature channels (\ie the CNN branch) leads to a significant performance drop across multiple datasets, further confirming the indispensability of payload information in improving traffic detection accuracy under complex network conditions.
Finally, when using only the multi-view feature extraction module (MFE) for traffic detection (\ie \model Only MFE), the model still achieves relatively competitive performance, further emphasizing the effectiveness of this module.

\subsection{Analysis on Hypergraph Construction} \label{sec: k}

In this work, we utilize Euclidean distance to measure the correlation between traffic flows and employ the KNN algorithm to construct a traffic hypergraph structure. To further analyze the impact of the hypergraph structure on the framework's performance, we conduct a parameter analysis for $K$. Specifically, we varies $K$ within the parameter space \{2, 3, 5, 10, 50, 100, 200\} and record the fluctuations in framework performance, as shown in Fig.~\ref{fig: topk}. It is evident that the performance exhibits a consistent fluctuation trend as $K$ changes. When the $K$ value is small (not exceeding 10), the framework's performance remains relatively stable; however, as the $K$ value increases, the performance shows a significant decline. This is because with a smaller $K$, traffic groups consist of only the most similar traffic instances, highlighting stronger higher-order correlations. When the $K$ value becomes larger, the traffic group gathers more traffic instances with varying degrees of similarity, significantly increasing the likelihood of introducing noisy flows, thereby affecting the framework's performance. Therefore, a relatively small $K$ value is favorable for capturing higher-order correlations between flows. In this paper, we set it to 3.

\subsection{Analysis on Dual-contrastive Constraints}\label{sec: Ablation}

We deploy ablation experiments to further analyze the impact of dual-contrastive constraints on the framework's performance. Specifically, we gradually remove one to two contrastive constraints from \model and observe the performance of the ablated models across five datasets, as shown in Table~\ref{tab: Abaltion_1} and \ref{tab: Abaltion_2}. From the results, we can see that removing some or all contrastive constraints leads to a certain degree of performance degradation in most cases, indicating that the contrastive self-supervised mechanism effectively enhances the framework's discriminative ability in scenarios with data imbalance and label scarcity. Moreover, using more contrastive constraints to optimize the framework generally results in greater performance improvements compared to using only a single contrastive constraint.

\begin{table*}[!htb] 
    \renewcommand\arraystretch{1.3} 
    \centering   
    \caption{Ablation analysis of different contrastive constraints on CIC-IOMT2024, UNSW-NB15 and Darknet2020.} 
    \resizebox{\textwidth}{!}{
\begin{tabular}{cc|cccc|cccc|cccc} 
\hline
\multicolumn{2}{c|}{Contrast}                 & \multicolumn{4}{c|}{CIC-IOMT2024}                                                                     & \multicolumn{4}{c|}{UNSW-NB15}                                                                               & \multicolumn{4}{c}{Darknet2020}                                                                                             \\ 
\hline
$\mathcal{L}_v$       & $\mathcal{L}_g$       & Accuracy                & Precison                & Recall                  & F1-score                & Accuracy                & Precison                       & Recall                  & F1-score                & Accuracy                & Precison                       & Recall                         & F1-score                        \\ 
\hline
-                     & -                     & 95.05$\pm$0.05          & \uline{93.73$\pm$0.34}  & 87.31$\pm$0.31          & 89.12$\pm$0.27          & 94.41$\pm$0.03          & 88.46$\pm$1.01                 & 83.27$\pm$0.52          & 85.13$\pm$0.53          & 94.30$\pm$0.11          & 84.77$\pm$1.77                 & 81.23$\pm$0.70                 & 80.75$\pm$0.82                  \\
\textbf{$\checkmark$} & -                     & \uline{95.27$\pm$0.02}  & 93.48$\pm$0.61          & \uline{88.77$\pm$0.12}  & \uline{90.21$\pm$0.13}  & \uline{94.48$\pm$0.07}  & 88.72$\pm$0.64                 & 82.77$\pm$0.37          & 85.11$\pm$0.36          & 94.50$\pm$0.05          & \uline{85.17}$\pm$\uline{0.46} & \uline{81.35}$\pm$\uline{0.49} & 81.53$\pm$0.46                  \\
-                     & \textbf{$\checkmark$} & 95.26$\pm$0.03          & 93.49$\pm$0.35          & 88.69$\pm$0.08          & 90.20$\pm$0.14          & 94.43$\pm$0.05          & \uline{88.87}$\pm$\uline{0.80} & \uline{83.42$\pm$0.51}  & \uline{85.29$\pm$0.44}  & \uline{94.50$\pm$0.05}  & 85.09$\pm$0.69                 & 81.13$\pm$0.42                 & \uline{81.59}$\pm$\uline{0.24}  \\
\textbf{$\checkmark$} & \textbf{$\checkmark$} & \textbf{95.27$\pm$0.01} & \textbf{93.75$\pm$0.46} & \textbf{88.89$\pm$0.12} & \textbf{90.36$\pm$0.10} & \textbf{94.49$\pm$0.03} & \textbf{89.78$\pm$0.21}        & \textbf{84.30$\pm$0.17} & \textbf{86.05$\pm$0.09} & \textbf{94.51$\pm$0.06} & \textbf{86.41$\pm$0.50}        & \textbf{82.62$\pm$0.48}        & \textbf{82.44$\pm$0.22}         \\
\hline
\end{tabular}}
\label{tab: Abaltion_1}
\end{table*}

\begin{table*}[!htb] 
    \renewcommand\arraystretch{1.3} 
    \centering     
    \caption{Ablation analysis of different contrastive constraints on ISCX-VPN2016 and USTC-TFC2016.}
    \resizebox{0.71\textwidth}{!}{
    \begin{tabular}{cc|cccc|cccc} 
\hline
\multicolumn{2}{c|}{Contrast}                 & \multicolumn{4}{c|}{ISCX-VPN2016}                                                                     & \multicolumn{4}{c}{USTC-TFC2016}                                                                       \\ 
\hline
$\mathcal{L}_v$       & $\mathcal{L}_g$       & Accuracy                & Precision               & Recall                  & F1-score                & Accuracy                & Precision               & Recall                  & F1-score                 \\ 
\hline
-                     & -                     & \uline{87.98$\pm$0.05}  & 90.25$\pm$1.11          & 90.06$\pm$0.18          & \uline{88.35$\pm$0.13}  & 97.59$\pm$0.03          & 96.92$\pm$0.11          & \uline{96.91$\pm$0.08}  & 96.84$\pm$0.08           \\
\textbf{$\checkmark$} & -                     & 87.95$\pm$0.08          & \uline{90.64$\pm$1.27}  & 90.09$\pm$0.17          & 88.26$\pm$0.20          & 97.61$\pm$0.02          & 96.90$\pm$0.06          & 96.88$\pm$0.09          & 96.78$\pm$0.03           \\
-                     & \textbf{$\checkmark$} & 87.95$\pm$0.09          & 90.25$\pm$0.87          & \uline{90.14$\pm$0.11}  & 88.32$\pm$0.18          & \uline{97.63$\pm$0.06}  & \uline{96.93$\pm$0.10}  & 96.90$\pm$0.04          & \uline{96.86$\pm$0.09}   \\
\textbf{$\checkmark$} & \textbf{$\checkmark$} & \textbf{88.04$\pm$0.05} & \textbf{91.44$\pm$0.46} & \textbf{90.25$\pm$0.12} & \textbf{88.36$\pm$0.10} & \textbf{97.67$\pm$0.04} & \textbf{96.98$\pm$0.10} & \textbf{96.99$\pm$0.08} & \textbf{96.90$\pm$0.06}  \\
\hline
\end{tabular}}
\label{tab: Abaltion_2}
\end{table*}

\begin{figure*}[!htb]
	\centering
	\includegraphics[width=\textwidth]{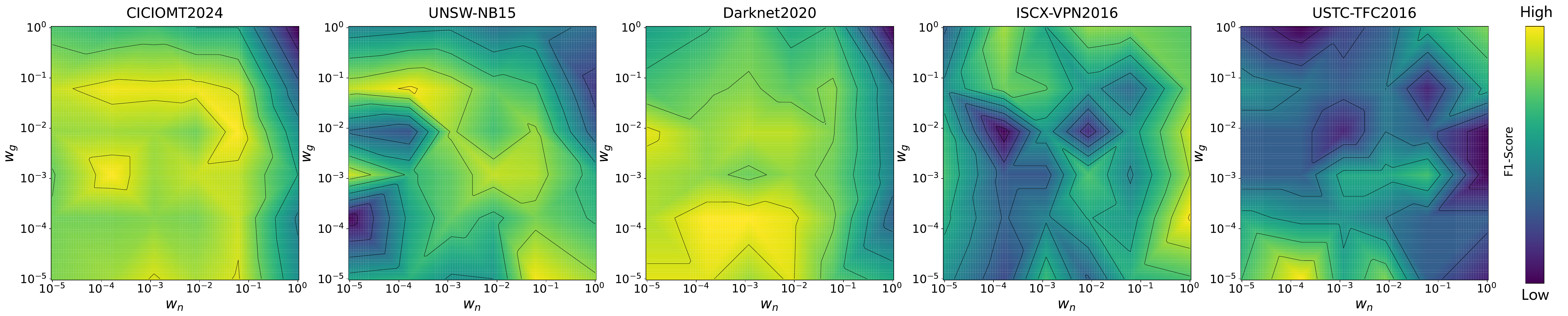}      
	\caption{The impact of different contrastive loss weights on framework performance.}
	\label{fig: CLW}
\end{figure*}


\begin{figure}[!htb]
	\centering
	\includegraphics[width=\linewidth]{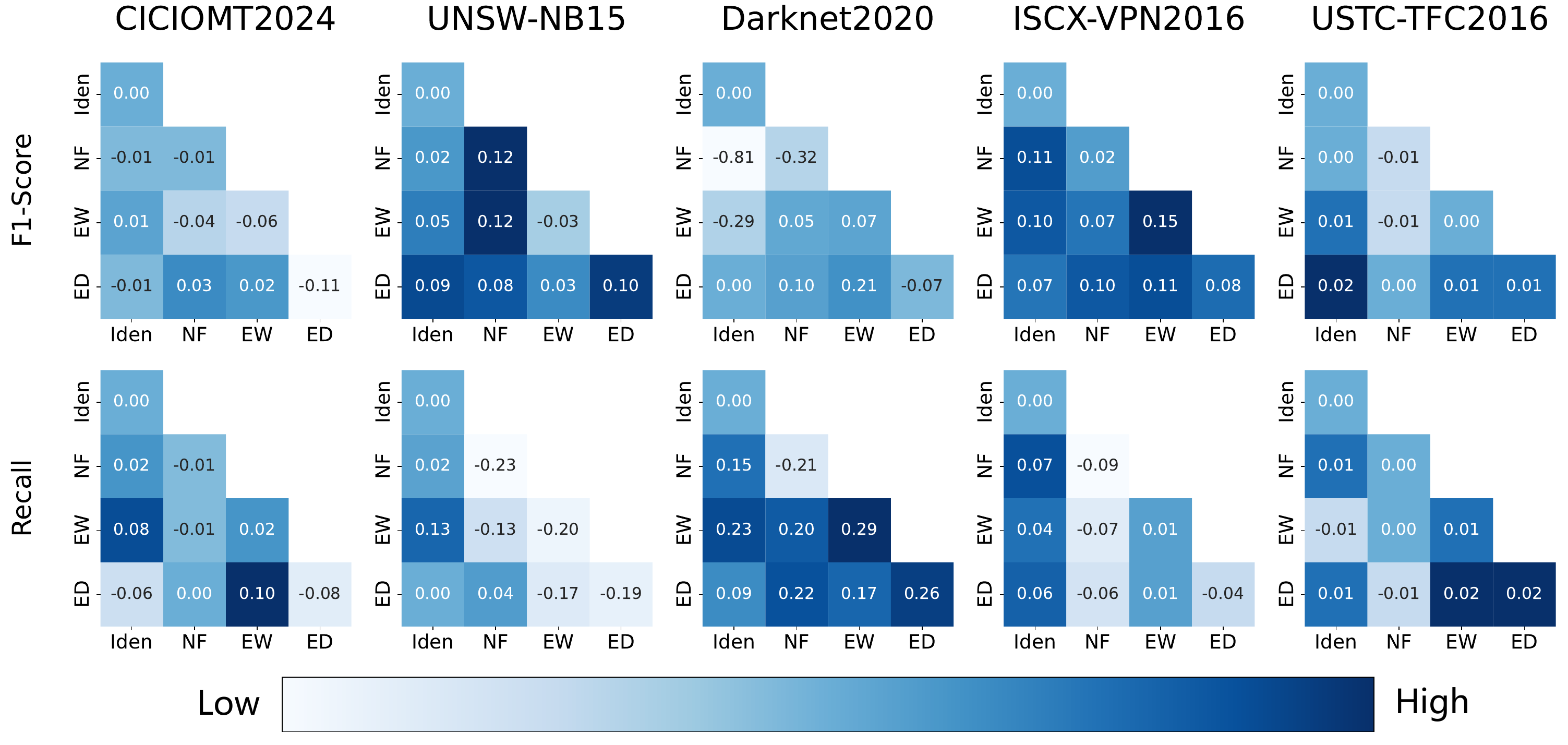}    
	\caption{Traffic Classifier gain (\%) when contrasting different augmentation pairs, compared to \model which stands for a no-augmentation version of our framework, under all datasets. ``Iden'' represents the original view.}
	\label{fig: aug}
\end{figure}

In order to deeply analyze the impact of contrastive loss weights on model performance, we conduct parametric experiments on two contrastive loss weight parameters, $\omega_n$ and $\omega_g$, using a grid search with a search range of $[10^{-5},... ,10^{0}]$. Due to the nonlinear parameter intervals, we use the quantile method to visualize our experimental results, and the experimental results are shown in Fig.~\ref{fig: CLW}.
The experimental results demonstrate that the weight parameters $\omega_n$ and $\omega_g$ of contrastive learning exhibit significant differences in their impact on model performance across different value ranges. When $\omega_n$ and $\omega_g$ are smaller than $10^{-4}$, the model shows relatively poor performance on most datasets but maintains a certain level of stability. This indicates that excessively small weights significantly diminish the contribution of contrastive learning to the model, without excessively degrading its overall performance. When $\omega_n$ and $\omega_g$ fall within the range of [$10^{-3},10^{-2}$], the model achieves peak performance on most datasets. However, compared to $\omega_g$, the model performance is more sensitive to variations in $\omega_n$, exhibiting more pronounced fluctuations. When $\omega_n$ and $\omega_g$ exceed $10^{-1}$, the model performance shows a significant decline, reaching its lowest point. This suggests that excessively high weights cause the model to overfit to specific contrastive learning features, thereby impairing its generalization ability. Based on the above analysis, we set the values of $\omega_n$ and $\omega_g$ within the range of [$10^{-3},10^{-2}$] to ensure that contrastive learning can stably enhance model performance.

\subsection{Analysis on Hypergraph Augmentation}\label{sec: Augmentation}



Different combinations of augmentation strategies may influence the effectiveness of contrastive learning. To evaluate the impact of data augmentation, we first compare the performance of an ablated model without augmentation (\ie \model w/o Aug) in the Table ~\ref{tab: Abaltion_mv1} and ~\ref{tab: Abaltion_mv_2}. The results show that the augmentation module provides consistent performance gains for the \model framework. We further analyze how different augmentation combinations affect model performance. In the experiments, we use a fixed masking rate of 0.4 across all strategies. As shown in Fig.~\ref{fig: aug}, several conclusions can be drawn:
1) Combining different augmentation methods generally improves model performance, particularly when distinct strategies are applied to each view, which is often more effective than using a single strategy;
2) Perturbing the hypergraph structure (\ie ED) typically yields stable positive gains, while perturbing hypergraph feature (especially NF) tends to degrade performance. This is primarily because structural perturbations introduce greater view diversity and variability, which helps the model learn more robust high-order correlations.

\section{Conclusion}\label{sec: conclusion}
In this paper, we propose a network traffic detection method called \model, which first employs a novel multi-view feature extractor to capture internal feature of traffic at a micro level. We then use the KNN algorithm to construct a traffic hypergraph and design two hypergraph data augmentation techniques, combined with bidirectional contrastive learning, to achieve precise network traffic detection. We evaluated the model's detection and generalization performance on five publicly available datasets. The experimental results demonstrate that: (1) The hypergraph-based contrastive learning method for network traffic detection outperforms existing approaches across multiple scenarios. (2) The multi-view feature learning method effectively captures the internal features of traffic. (3) Both hypergraph data augmentation and hypergraph contrastive learning contribute to improving the model's detection and generalization capabilities.

\bibliographystyle{IEEEtran} 
\bibliography{mybib,IEEEabrv}

@article{zhang2014robust,
  title={Robust network traffic classification},
  author={Zhang, Jun and Chen, Xiao and Xiang, Yang and Zhou, Wanlei and Wu, Jie},
  journal={IEEE/ACM transactions on networking},
  volume={23},
  number={4},
  pages={1257--1270},
  year={2014},
  publisher={IEEE}
}

@inproceedings{kotpalliwar2015classification,
  title={Classification of attacks using support vector machine (svm) on kddcup'99 ids database},
  author={Kotpalliwar, Manjiri V and Wajgi, Rakhi},
  booktitle={2015 Fifth International Conference on Communication Systems and Network Technologies},
  pages={987--990},
  year={2015},
  organization={IEEE}
}

@inproceedings{kokila2014ddos,
  title={DDoS detection and analysis in SDN-based environment using support vector machine classifier},
  author={Kokila, RT and Selvi, S Thamarai and Govindarajan, Kannan},
  booktitle={2014 sixth international conference on advanced computing (ICoAC)},
  pages={205--210},
  year={2014},
  organization={IEEE}
}

@article{kwon2019survey,
  title={A survey of deep learning-based network anomaly detection},
  author={Kwon, Donghwoon and Kim, Hyunjoo and Kim, Jinoh and Suh, Sang C and Kim, Ikkyun and Kim, Kuinam J},
  journal={Cluster Computing},
  volume={22},
  number={Suppl 1},
  pages={949--961},
  year={2019},
  publisher={Springer}
}

@article{paxson1994empirically,
  title={Empirically derived analytic models of wide-area TCP connections},
  author={Paxson, Vern},
  journal={IEEE/ACM transactions on Networking},
  volume={2},
  number={4},
  pages={316--336},
  year={1994},
  publisher={IEEE}
}

@inproceedings{lang2003synthetic,
  title={A synthetic traffic model for Half-Life},
  author={Lang, Tanja and Armitage, Grenville and Branch, Phillip and Choo, Hwan-Yi},
  booktitle={Australian Telecommunications Networks \& Applications Conference},
  volume={2003},
  year={2003},
  pages={1--5}
}

@article{nguyen2008survey,
  title={A survey of techniques for internet traffic classification using machine learning},
  author={Nguyen, Thuy TT and Armitage, Grenville},
  journal={IEEE communications surveys \& tutorials},
  volume={10},
  number={4},
  pages={56--76},
  year={2008},
  publisher={IEEE}
}

@article{liu2019machine,
  title={Machine learning and deep learning methods for intrusion detection systems: A survey},
  author={Liu, Hongyu and Lang, Bo},
  journal={applied sciences},
  volume={9},
  number={20},
  pages={4396},
  year={2019},
  publisher={mdpi}
}

@article{dadkhah2024ciciomt2024,
  title={CICIoMT2024: Attack Vectors in Healthcare devices-A Multi-Protocol Dataset for Assessing IoMT Device Security},
  author={Dadkhah, Sajjad and Carlos Pinto Neto, Euclides and Ferreira, Raphael and Chukwuka Molokwu, Reginald and Sadeghi, Somayeh and Ghorbani, Ali},
  journal={Internet of Things},
  year={2024},
  volume={28},
  pages={101351}
}

@inproceedings{moustafa2015unsw,
  title={UNSW-NB15: a comprehensive data set for network intrusion detection systems (UNSW-NB15 network data set)},
  author={Moustafa, Nour and Slay, Jill},
  booktitle={2015 military communications and information systems conference (MilCIS)},
  pages={1--6},
  year={2015},
  organization={IEEE}
}

@inproceedings{habibi2020didarknet,
  title={Didarknet: A contemporary approach to detect and characterize the darknet traffic using deep image learning},
  author={Habibi Lashkari, Arash and Kaur, Gurdip and Rahali, Abir},
  booktitle={Proceedings of the 2020 10th International Conference on Communication and Network Security},
  pages={1--13},
  year={2020}
}

@inproceedings{draper2016characterization,
  title={Characterization of encrypted and vpn traffic using time-related},
  author={Draper-Gil, Gerard and Lashkari, Arash Habibi and Mamun, Mohammad Saiful Islam and Ghorbani, Ali A},
  booktitle={Proceedings of the 2nd international conference on information systems security and privacy (ICISSP)},
  pages={407--414},
  year={2016}
}

@inproceedings{li2022graphddos,
  title={Graphddos: Effective ddos attack detection using graph neural networks},
  author={Li, Yuzhen and Li, Renjie and Zhou, Zhou and Guo, Jiang and Yang, Wei and Du, Meijie and Liu, Qingyun},
  booktitle={2022 IEEE 25th International Conference on Computer Supported Cooperative Work in Design (CSCWD)},
  pages={1275--1280},
  year={2022},
  organization={IEEE}
}

@inproceedings{wang2020app,
  title={App-net: A hybrid neural network for encrypted mobile traffic classification},
  author={Wang, Xin and Chen, Shuhui and Su, Jinshu},
  booktitle={IEEE INFOCOM 2020-IEEE Conference on Computer Communications Workshops (INFOCOM WKSHPS)},
  pages={424--429},
  year={2020},
  organization={IEEE}
}

@article{shen2021accurate,
  title={Accurate decentralized application identification via encrypted traffic analysis using graph neural networks},
  author={Shen, Meng and Zhang, Jinpeng and Zhu, Liehuang and Xu, Ke and Du, Xiaojiang},
  journal={IEEE Transactions on Information Forensics and Security},
  volume={16},
  pages={2367--2380},
  year={2021},
  publisher={IEEE}
}

@inproceedings{kipf2016semi,
  title={Semi-Supervised Classification with Graph Convolutional Networks},
  author={Kipf, Thomas N and Welling, Max},
  booktitle={International Conference on Learning Representations},
  year={2017},
  pages={1--14}
}

@inproceedings{feng2019hypergraph,
  title={Hypergraph neural networks},
  author={Feng, Yifan and You, Haoxuan and Zhang, Zizhao and Ji, Rongrong and Gao, Yue},
  booktitle={Proceedings of the AAAI conference on artificial intelligence},
  volume={33},
  number={01},
  pages={3558--3565},
  year={2019}
}

@article{gao2022hgnn+,
  title={HGNN+: General hypergraph neural networks},
  author={Gao, Yue and Feng, Yifan and Ji, Shuyi and Ji, Rongrong},
  journal={IEEE Transactions on Pattern Analysis and Machine Intelligence},
  volume={45},
  number={3},
  pages={3181--3199},
  year={2022},
  publisher={IEEE}
}

@article{yadati2019hypergcn,
  title={Hypergcn: A new method for training graph convolutional networks on hypergraphs},
  author={Yadati, Naganand and Nimishakavi, Madhav and Yadav, Prateek and Nitin, Vikram and Louis, Anand and Talukdar, Partha},
  journal={Advances in neural information processing systems},
  volume={32},
  year={2019},
  pages={1--12}
}

@inproceedings{moore2005toward,
  title={Toward the Accurate Identification of Network Applications},
  author={Moore, Andrew W and Papagiannaki, Konstantina},
  booktitle={Passive and Active Network Measurement: 6th International Workshop, PAM 2005, Boston, MA, USA, March 31-April 1, 2005, Proceedings},
  volume={3431},
  pages={41--54},
  year={2005},
  organization={Springer}
}

@inproceedings{taylor2016appscanner,
  title={Appscanner: Automatic fingerprinting of smartphone apps from encrypted network traffic},
  author={Taylor, Vincent F and Spolaor, Riccardo and Conti, Mauro and Martinovic, Ivan},
  booktitle={2016 IEEE European Symposium on Security and Privacy (EuroS\&P)},
  pages={439--454},
  year={2016},
  organization={IEEE}
}

@inproceedings{van2020flowprint,
  title={Flowprint: Semi-supervised mobile-app fingerprinting on encrypted network traffic},
  author={Van Ede, Thijs and Bortolameotti, Riccardo and Continella, Andrea and Ren, Jingjing and Dubois, Daniel J and Lindorfer, Martina and Choffnes, David and Van Steen, Maarten and Peter, Andreas},
  booktitle={Network and distributed system security symposium (NDSS)},
  volume={27},
  year={2020},
  pages={1--18}
}

@inproceedings{liu2019fs,
  title={Fs-net: A flow sequence network for encrypted traffic classification},
  author={Liu, Chang and He, Longtao and Xiong, Gang and Cao, Zigang and Li, Zhen},
  booktitle={IEEE INFOCOM 2019-IEEE Conference On Computer Communications},
  pages={1171--1179},
  year={2019},
  organization={IEEE}
}

@article{zheng2022mtt,
  title={MTT: an efficient model for encrypted network traffic classification using multi-task transformer},
  author={Zheng, Weiping and Zhong, Jianhao and Zhang, Qizhi and Zhao, Gansen},
  journal={Applied Intelligence},
  volume={52},
  number={9},
  pages={10741--10756},
  year={2022},
  publisher={Springer}
}

@article{wang2024netmamba,
  title={Netmamba: Efficient network traffic classification via pre-training unidirectional mamba},
  author={Wang, Tongze and Xie, Xiaohui and Wang, Wenduo and Wang, Chuyi and Zhao, Youjian and Cui, Yong},
  booktitle={2024 IEEE 32nd International Conference on Network Protocols (ICNP)},
  pages={1--11},
  year={2024},
  organization={IEEE}
}

@inproceedings{wang2017malware,
  title={Malware traffic classification using convolutional neural network for representation learning},
  author={Wang, Wei and Zhu, Ming and Zeng, Xuewen and Ye, Xiaozhou and Sheng, Yiqiang},
  booktitle={2017 International conference on information networking (ICOIN)},
  pages={712--717},
  year={2017},
  organization={IEEE}
}

@inproceedings{saber2018encrypted,
  title={Encrypted traffic classification: Combining over-and under-sampling through a pca-svm},
  author={Saber, Abid and Fergani, Belkacem and Abbas, Moncef},
  booktitle={2018 3rd International Conference on Pattern Analysis and Intelligent Systems (PAIS)},
  pages={1--5},
  year={2018},
  organization={IEEE}
}

@article{huoh2022flow,
  title={Flow-based encrypted network traffic classification with graph neural networks},
  author={Huoh, Ting-Li and Luo, Yan and Li, Peilong and Zhang, Tong},
  journal={IEEE Transactions on Network and Service Management},
  volume={20},
  number={2},
  pages={1224--1237},
  year={2022},
  publisher={IEEE}
}

@inproceedings{barsellotti2023ftg,
  title={FTG-Net: Hierarchical flow-to-traffic graph neural network for DDoS attack detection},
  author={Barsellotti, Luca and De Marinis, Lorenzo and Cugini, Filippo and Paolucci, Francesco},
  booktitle={2023 IEEE 24th International Conference on High Performance Switching and Routing (HPSR)},
  pages={173--178},
  year={2023},
  organization={IEEE}
}

@article{han2024intrusion,
  title={Intrusion Detection for Encrypted Flows Using Single Feature Based on Graph Integration Theory},
  author={Han, Ying and Wang, Xinlei and He, Mingshu and Wang, Xiaojuan and Guo, Shize},
  journal={IEEE Internet of Things Journal},
  volume={11},
  number={10},
  pages={17589--17601},
  year={2024},
  publisher={IEEE}
}

@article{han2024gnn,
  title={DE-GNN: Dual embedding with graph neural network for fine-grained encrypted traffic classification},
  author={Han, Xinbo and Xu, Guizhong and Zhang, Meng and Yang, Zheng and Yu, Ziyang and Huang, Weiqing and Meng, Chen},
  journal={Computer Networks},
  volume={245},
  pages={110372},
  year={2024},
  publisher={Elsevier}
}

@article{yang2024hrnn,
  title={HRNN: Hypergraph Recurrent Neural Network for Network Intrusion Detection},
  author={Yang, Zhe and Ma, Zitong and Zhao, Wenbo and Li, Lingzhi and Gu, Fei},
  journal={Journal of Grid Computing},
  volume={22},
  number={2},
  pages={1--15},
  year={2024},
  publisher={Springer}
}

@article{jin2022heterogeneous,
  title={Heterogeneous feature augmentation for ponzi detection in ethereum},
  author={Jin, Chengxiang and Jin, Jie and Zhou, Jiajun and Wu, Jiajing and Xuan, Qi},
  journal={IEEE Transactions on Circuits and Systems II: Express Briefs},
  volume={69},
  number={9},
  pages={3919--3923},
  year={2022},
  publisher={IEEE}
}

@inproceedings{ying2018graph,
  title={Graph convolutional neural networks for web-scale recommender systems},
  author={Ying, Rex and He, Ruining and Chen, Kaifeng and Eksombatchai, Pong and Hamilton, William L and Leskovec, Jure},
  booktitle={Proceedings of the 24th ACM SIGKDD international conference on knowledge discovery \& data mining},
  pages={974--983},
  year={2018}
}

@inproceedings{chen2020hypergraph,
  title={Hypergraph attention networks},
  author={Chen, Chaofan and Cheng, Zelei and Li, Zuotian and Wang, Manyi},
  booktitle={2020 IEEE 19th International Conference on Trust, Security and Privacy in Computing and Communications (TrustCom)},
  pages={1560--1565},
  year={2020},
  organization={IEEE}
}

@inproceedings{wang2020next,
  title={Next-item recommendation with sequential hypergraphs},
  author={Wang, Jianling and Ding, Kaize and Hong, Liangjie and Liu, Huan and Caverlee, James},
  booktitle={Proceedings of the 43rd international ACM SIGIR conference on research and development in information retrieval},
  pages={1101--1110},
  year={2020}
}

@article{yu2023routing,
  title={Routing hypergraph convolutional recurrent network for network traffic prediction},
  author={Yu, Weihao and Ruan, Ke and Tang, Hong and Huang, Jin},
  journal={Applied Intelligence},
  volume={53},
  number={12},
  pages={16126--16137},
  year={2023},
  publisher={Springer}
}

@inproceedings{pu2011hypergraph,
  title={Hypergraph clustering for better network traffic inspection},
  author={Pu, Li and Faltings, Boi},
  booktitle={Working Notes for the 2011 IJCAI Workshop on Intelligent Security (SecArt)},
  pages={18--25},
  year={2011}
}

@inproceedings{bikmukhamedov2021multi,
  title={Multi-class network traffic generators and classifiers based on neural networks},
  author={Bikmukhamedov, RF and Nadeev, AF},
  booktitle={2021 Systems of Signals Generating and Processing in the Field of on Board Communications},
  pages={1--7},
  year={2021},
  organization={IEEE}
}

@article{zhou2022behavior,
  title={Behavior-aware account de-anonymization on ethereum interaction graph},
  author={Zhou, Jiajun and Hu, Chenkai and Chi, Jianlei and Wu, Jiajing and Shen, Meng and Xuan, Qi},
  journal={IEEE Transactions on Information Forensics and Security},
  volume={17},
  pages={3433--3448},
  year={2022},
  publisher={IEEE}
}

@inproceedings{zhou2020data-cikm,
  title={Data augmentation for graph classification},
  author={Zhou, Jiajun and Shen, Jie and Xuan, Qi},
  booktitle={Proceedings of the 29th ACM International Conference on Information \& Knowledge Management},
  pages={2341--2344},
  year={2020}
}

@article{Mevolve,
  title={M-evolve: structural-mapping-based data augmentation for graph classification},
  author={Zhou, Jiajun and Shen, Jie and Yu, Shanqing and Chen, Guanrong and Xuan, Qi},
  journal={IEEE Transactions on Network Science and Engineering},
  volume={8},
  number={1},
  pages={190--200},
  year={2020},
  publisher={IEEE}
}

@article{zhou2022data,
  title={Data augmentation on graphs: A technical survey},
  author={Zhou, Jiajun and Xie, Chenxuan and Gong, Shengbo and Wen, Zhenyu and Zhao, Xiangyu and Xuan, Qi and Yang, Xiaoniu},
  journal={ACM Computing Surveys},
  volume={57},
  number={11},
  pages={1--34},
  year={2025},
  publisher={ACM New York, NY}
}

@article{zhou2021robustecd,
  title={RobustECD: Enhancement of network structure for robust community detection},
  author={Zhou, Jiajun and Chen, Zhi and Du, Min and Chen, Lihong and Yu, Shanqing and Chen, Guanrong and Xuan, Qi},
  journal={IEEE Transactions on Knowledge and Data Engineering},
  volume={35},
  number={1},
  pages={842--856},
  year={2021},
  publisher={IEEE}
}

@article{lstm,
  title={Long short-term memory},
  author={Hochreiter, Sepp and Schmidhuber, J{\"u}rgen},
  journal={Neural computation},
  volume={9},
  number={8},
  pages={1735--1780},
  year={1997},
  publisher={MIT press}
}

@article{hwang2019lstm,
  title={An LSTM-based deep learning approach for classifying malicious traffic at the packet level},
  author={Hwang, Ren-Hung and Peng, Min-Chun and Nguyen, Van-Linh and Chang, Yu-Lun},
  journal={Applied Sciences},
  volume={9},
  number={16},
  pages={3414},
  year={2019},
  publisher={MDPI}
}

@article{hu2023tcgnn,
  title={TCGNN: Packet-grained network traffic classification via Graph Neural Networks},
  author={Hu, Guangwu and Xiao, Xi and Shen, Meng and Zhang, Bin and Yan, Xia and Liu, Yunxia},
  journal={Engineering Applications of Artificial Intelligence},
  volume={123},
  pages={106531},
  year={2023},
  publisher={Elsevier}
}

@inproceedings{zhang2019framework,
  title={A framework for resource-aware online traffic classification using CNN},
  author={Zhang, Wanqian and Wang, Junxiao and Chen, Sheng and Qi, Heng and Li, Keqiu},
  booktitle={Proceedings of the 14th International Conference on Future Internet Technologies},
  pages={1--6},
  year={2019}
}

@article{jin2024time,
  title={Time-aware metapath feature augmentation for ponzi detection in ethereum},
  author={Jin, Chengxiang and Zhou, Jiajun and Jin, Jie and Wu, Jiajing and Xuan, Qi},
  journal={IEEE Transactions on Network Science and Engineering},
  volume={11},
  number={4},
  pages={3747--3758},
  year={2024},
  publisher={IEEE}
}

@inproceedings{chen2019supervised,
  title={Supervised Community Detection with Line Graph Neural Networks},
  author={Chen, Zhengdao and Li, Lisha and Bruna, Joan},
  booktitle={International conference on learning representations},
  year={2020},
  pages={1--23}
}

@inproceedings{TFE-GNN,
  title={Tfe-gnn: A temporal fusion encoder using graph neural networks for fine-grained encrypted traffic classification},
  author={Zhang, Haozhen and Yu, Le and Xiao, Xi and Li, Qing and Mercaldo, Francesco and Luo, Xiapu and Liu, Qixu},
  booktitle={Proceedings of the ACM Web Conference 2023},
  pages={2066--2075},
  year={2023}
}

@article{zhou2025ncgnn,
  title={Clarify Confused Nodes via Separated Learning},
  author={Zhou, Jiajun and Gong, Shengbo and Chen, Xuanze and Xie, Chenxuan and Yu, Shanqing and Xuan, Qi and Yang, Xiaoniu},
  journal={IEEE Transactions on Pattern Analysis and Machine Intelligence},
  year={2025},
  volume={47},
  number={4},
  pages={2882-2896}
}

@article{shen2025robust,
  title={Robust Detection of Malicious Encrypted Traffic via Contrastive Learning},
  author={Shen, Meng and Wu, Jinhe and Ye, Ke and Xu, Ke and Xiong, Gang and Zhu, Liehuang},
  journal={IEEE Transactions on Information Forensics and Security},
  year={2025},
  volume={20},
  number={},
  pages={4228-4242},
  publisher={IEEE}
}

@article{zhang2024enhanced,
  title={Enhanced few-shot malware traffic classification via integrating knowledge transfer with neural architecture search},
  author={Zhang, Xixi and Wang, Qin and Qin, Maoyang and Wang, Yu and Ohtsuki, Tomoaki and Adebisi, Bamidele and Sari, Hikmet and Gui, Guan},
  journal={IEEE Transactions on Information Forensics and Security},
  year={2024},
  volume={19},
  pages={5245--5256},
  publisher={IEEE}
}

@article{liu2024atvitsc,
  title={ATVITSC: A novel encrypted traffic classification method based on deep learning},
  author={Liu, Ya and Wang, Xiao and Qu, Bo and Zhao, Fengyu},
  journal={IEEE Transactions on Information Forensics and Security},
  year={2024},
  volume={19},
  number={},
  pages={9374-9389},
  publisher={IEEE}
}

@article{du2023breaking,
  title={Breaking the anonymity of ethereum mixing services using graph feature learning},
  author={Du, Hanbiao and Che, Zheng and Shen, Meng and Zhu, Liehuang and Hu, Jiankun},
  journal={IEEE Transactions on Information Forensics and Security},
  volume={19},
  pages={616--631},
  year={2023},
  publisher={IEEE}
}

@article{che2024across,
  author={Che, Zheng and Shen, Meng and Tan, Zhehui and Du, Hanbiao and Wang, Wei and Chen, Ting and Zhao, Qinglin and Xie, Yong and Zhu, Liehuang},
  journal={IEEE Transactions on Information Forensics and Security}, 
  title={Across-Platform Detection of Malicious Cryptocurrency Accounts via Interaction Feature Learning}, 
  year={2025},
  volume={20},
  number={},
  pages={4783-4798}
}

\begin{IEEEbiography}[{\includegraphics[width=1in,height=1.25in,clip,keepaspectratio]{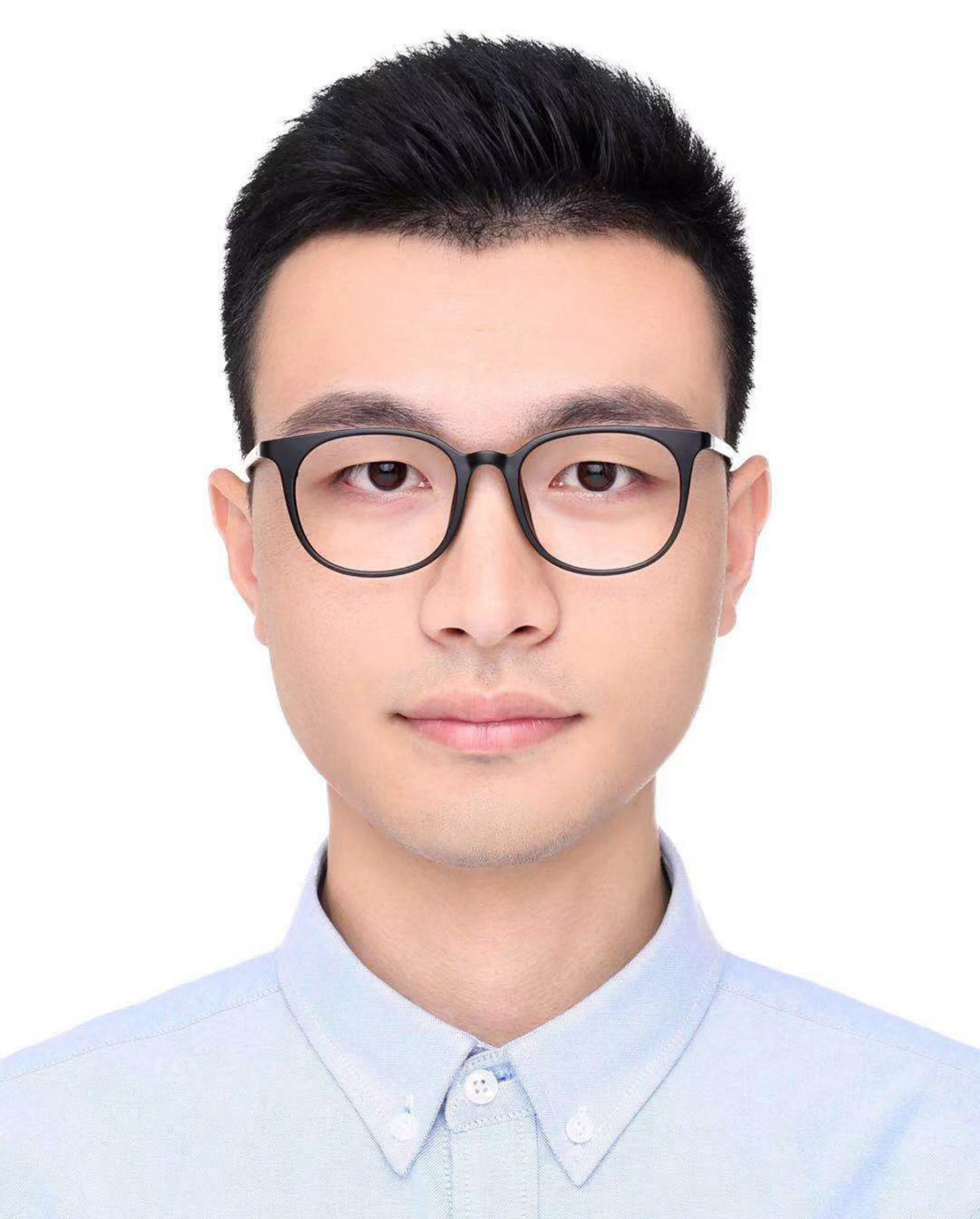}}]{Jiajun Zhou}
	received the Ph.D degree in control theory and engineering from Zhejiang University of Technology, Hangzhou, China, in 2023. He is currently a Postdoctoral Research Fellow with the Institute of Cyberspace Security, Zhejiang University of Technology. His current research interests include graph data mining, cyberspace security and data management.
\end{IEEEbiography}
\vspace{-35pt}

\begin{IEEEbiography}[{\includegraphics[width=1in,height=1.25in,clip,keepaspectratio]{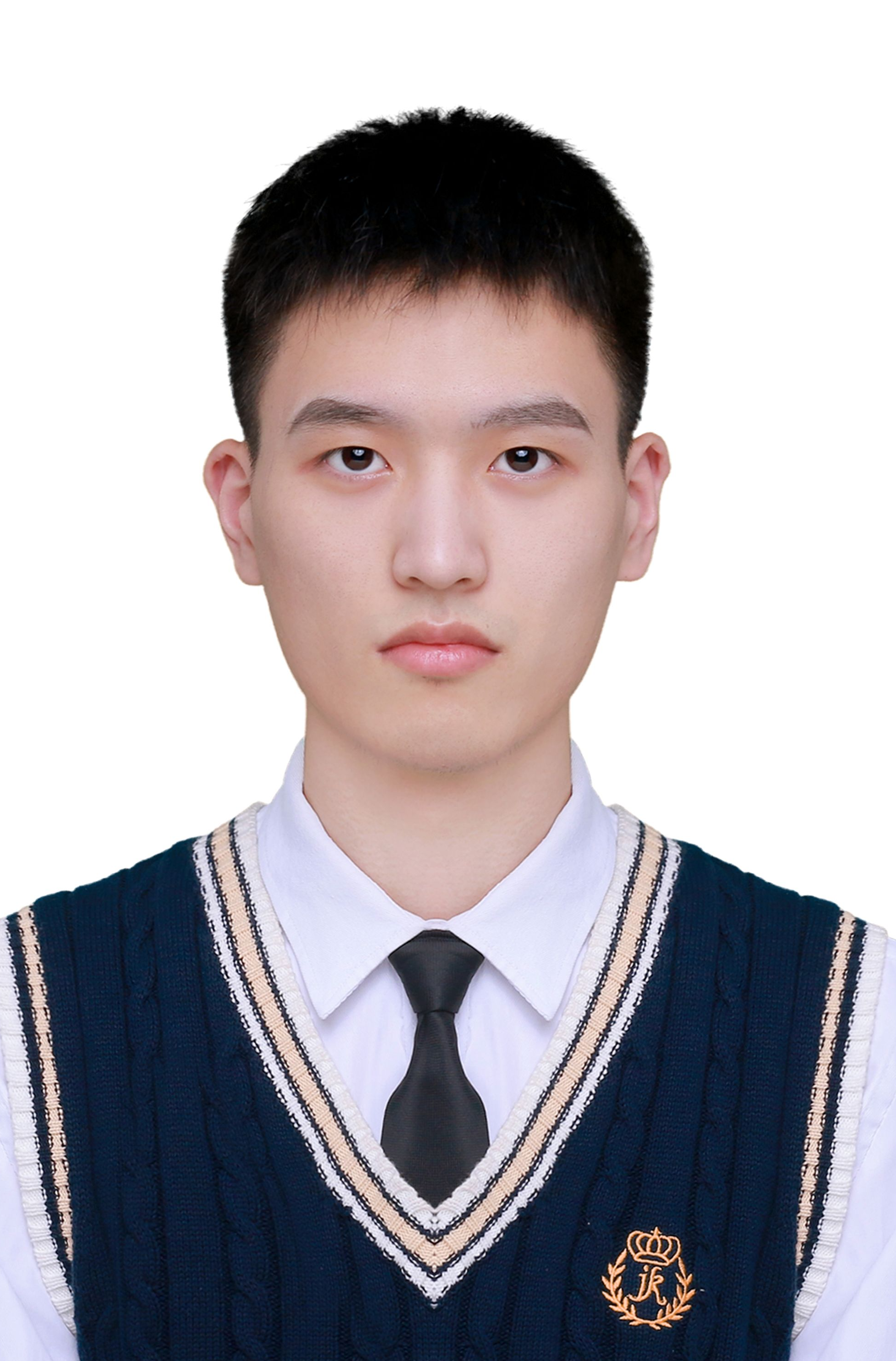}}]{Wentao Fu}
	received the B.S. degree in Automation from Xi'an University of Posts and Telecommunications, Xi'an, China, in 2023. He is currently pursuing his master's degree at the Institute of Cyberspace Security, Zhejiang University of Technology, China. His current research interests include cyberspace security and network intrusion detection.
\end{IEEEbiography}
\vspace{-35pt}

\begin{IEEEbiography}[{\includegraphics[width=1in,height=1.25in,clip,keepaspectratio]{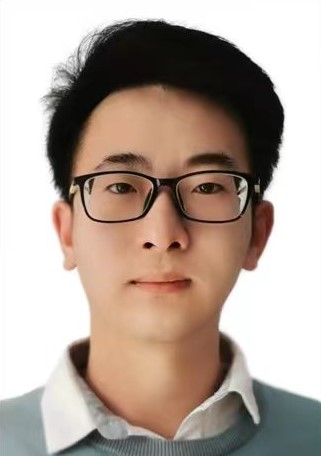}}]{Hao Song}
	received the B.S. degree in Communication Engineering from Nanjing Institute of Technology, Nanjing, China, in 2021. He is currently pursuing his master's degree at the Institute of Cyberspace Security, Zhejiang University of Technology, China. His current research interests include cyberspace security and network intrusion detection.
\end{IEEEbiography}
\vspace{-35pt}

\begin{IEEEbiography}[{\includegraphics[width=1in,height=1.25in,clip,keepaspectratio]{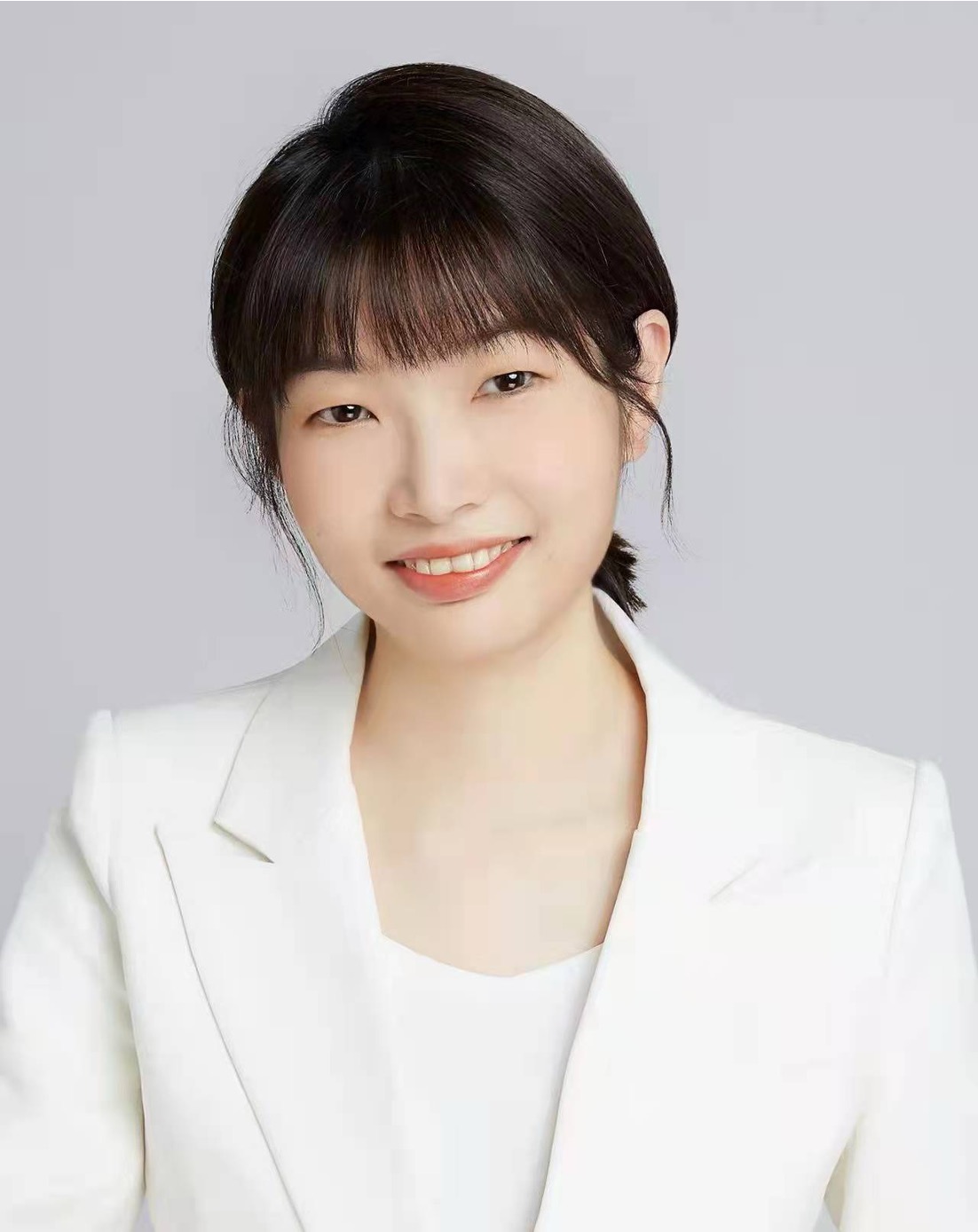}}]{Shanqing Yu}
	received the M.S. degree from the School of Computer Engineering and Science, Shanghai University, China, in 2008 and received the M.S. degree from the Graduate School of Information, Production and Systems, Waseda University, Japan, in 2008, and the Ph.D. degree, in 2011, respectively. She is currently a Lecturer at the Institute of Cyberspace Security and the College of Information Engineering, Zhejiang University of Technology, Hangzhou, China. Her research interests cover intelligent computation and data mining.
\end{IEEEbiography}
\vspace{-35pt}

\begin{IEEEbiography}[{\includegraphics[width=1in,height=1.25in,clip,keepaspectratio]{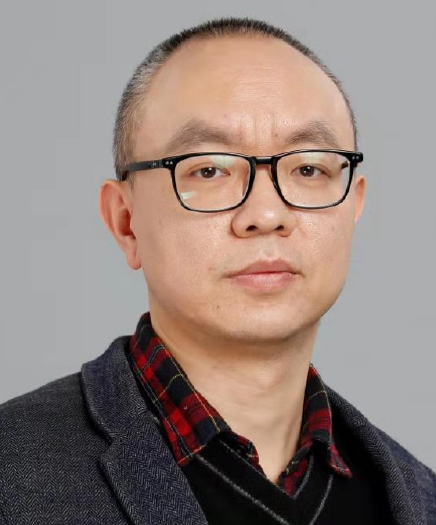}}]{Qi Xuan}(M'18) received the BS and PhD degrees in control theory and engineering from Zhejiang University, Hangzhou, China, in 2003 and 2008, respectively. He was a Post-Doctoral Researcher with the Department of Information Science and Electronic Engineering, Zhejiang University, from 2008 to 2010, respectively, and a Research Assistant with the Department of Electronic Engineering, City University of Hong Kong, Hong Kong, in 2010 and 2017. From 2012 to 2014, he was a Post-Doctoral Fellow with the Department of Computer Science, University of California at Davis, CA, USA. He is a senior member of the IEEE and is currently a Professor with the Institute of Cyberspace Security, College of Information Engineering, Zhejiang University of Technology, Hangzhou, China. His current research interests include network science, graph data mining, cyberspace security, machine learning, and computer vision.
\end{IEEEbiography}

\end{document}